\documentclass[useAMS,usenatbib]{mn2e}
\usepackage{epsfig}

\def\mnras{MNRAS}  % Monthly Notices of the RAS
\def\apj{ApJ}      % The Astrophysical Journal
\def\apjl{ApJL}    % The Astrophysical Journal Letters
\def\apjs{ApJS}    % The Astrophysical Journal Supplement
\def\aap{A\&A}     % Astronomy and Astrophysics
\def\aj{AJ}        % The Astronomical Journal
\def\araa{ARAA}    % Annual Reviews of Astronomy & Astrophysics
\def\nat{Nature}   % Nature
\def\gadget{{\small GADGET-2} }
\def\flash{{\small FLASH} }
\def\enzo{{\small ENZO} }
\def\art{{\small ART} }
\def\gasoline{{\small GASOLINE} }

\title[Entropy Cores in Clusters]{
On the Origin of Cores in Simulated Galaxy Clusters
}
\author[N. L. Mitchell et al.]{N. L. Mitchell$^1$\thanks{E-mail:
n.l.mitchell@durham.ac.uk}, I. G. McCarthy$^{1,2}$, R. G. Bower$^1$, 
T. Theuns$^{1,3}$, R. A. Crain$^1$
\\
\\
$^1$Department of Physics, Durham University, South Road,
Durham, DH1 3LE\\
$^2$Kavli Institute for Cosmology, University of Cambridge, Madingley Road,
Cambridge, CB3 0HA\\
$^3$Department of Physics, University of Antwerp, Campus
Groenenborger, Groenenborgerlaan 171, B-2020 Antwerp, Belgium
}

\begin{document}

\date{Accepted 2009 January 21.  Received 2009 January 21; in original form 2008 August 26}

\pagerange{\pageref{firstpage}--\pageref{lastpage}} \pubyear{2008}

\maketitle

\label{firstpage}

\begin{abstract}

The diffuse plasma that fills galaxy groups and clusters 
(the intracluster medium, hereafter ICM) is a by-product of 
galaxy formation. The present thermal state of this gas 
results from a competition between gas cooling and heating. 
The heating comes from two distinct sources: gravitational 
heating associated with the collapse of the dark matter halo 
and additional thermal input from the formation of galaxies 
and their black holes.  A long term goal of this research 
is to decode the observed temperature, density and entropy 
profiles of clusters and to understand the relative roles of 
these processes. However, a long standing problem has been that 
cosmological simulations based on smoothed particle hydrodynamics 
(SPH) and Eulerian mesh-based codes predict different results even 
when cooling and galaxy/black hole heating are switched 
off.  Clusters formed in SPH simulations show near powerlaw 
entropy profiles, while those formed in Eulerian simulations 
develop a core and do not allow gas to reach such low 
entropies. Since the cooling rate is closely connected to 
the minimum entropy of the gas distribution, the differences 
are of potentially key importance.

In this paper, we investigate the origin of this discrepancy. 
By comparing simulations run using the \gadget SPH code and 
the \flash adaptive Eulerian mesh code, we show that the 
discrepancy arises during the idealised merger of two
clusters, and that the differences are not the result of the 
lower effective resolution of Eulerian cosmological 
simulations.  The difference is not sensitive to 
the minimum mesh size (in Eulerian codes) or the number of 
particles used (in SPH codes).  We investigate whether the 
difference is the result of the different gravity solvers, 
the Galilean non-invariance of the mesh code or an effect of 
unsuitable artificial viscosity in the SPH code.  Instead, 
we find that the difference is inherent to the treatment of 
vortices in the two codes.  Particles in the SPH
simulations retain a close connection to their initial 
entropy, while this connection is much weaker in the mesh 
simulations.  The origin of this difference lies 
in the treatment of eddies and fluid instabilities. These are 
suppressed in the SPH simulations, while the cluster mergers 
generate strong vortices in the Eulerian simulations that 
very efficiently mix the fluid and erase the low entropy gas.  
We discuss the potentially profound implications of 
these results.
\end{abstract}

\begin{keywords}
hydrodynamics --- methods: N-body simulations --- galaxies: 
clusters: general --- cosmology: theory
\end{keywords}

\section{Introduction}

There has been a great deal of attention devoted in recent 
years to the properties of the hot X-ray-emitting plasma 
(the intracluster medium, hereafter ICM) in the central 
regions of massive galaxy groups and clusters.  To a large 
extent, the focus has been on the competition between 
radiative cooling losses and various mechanisms that could be 
heating the gas and therefore (at least partially) offsetting 
the effects of cooling.  Prior to the launch of the {\it 
Chandra} and {\it XMM-Newton} X-ray observatories, 
it was generally accepted that large quantities of the ICM 
should be cooling down to very low energies, where it would 
cease to emit X-rays and eventually condense out 
into molecular clouds and form stars (Fabian 1994).  
However, the amount of cold gas and star formation actually 
observed in most systems is well below what was expected 
based on analyses of {\it ROSAT}, {\it ASCA}, and {\it 
Einstein} X-ray data (e.g., Voit \& Donahue 1995; Edge 2001).  
New high spatial and spectral resolution data from {\it 
Chandra} and {\it XMM-Newton} has shown that, while cooling 
is clearly important in many groups and clusters (so-called 
`cool core' systems), relatively little gas is actually 
cooling out of the X-ray band to temperatures below 
$\sim10^7$ K (Peterson et al.\ 2003).  

It is now widely believed that some energetic form of 
non-gravitational heating is compensating for the 
losses due to cooling. Indeed, it seems likely that such 
heating goes beyond a simple compensation for the radiated
energy. As well as having a profound effect on the properties
of the ICM, it seems that the heat input also has 
important consequences for the formation and evolution of the 
central brightest cluster galaxy (BCG) and therefore for the 
bright end of the galaxy luminosity function (e.g., 
Benson et al.\ 2003; Bower et al.\ 2006, 2008).  The thermal 
state of the ICM therefore provides an important probe of 
these processes and a concerted
theoretical effort has been undertaken to explore the 
effects of various heating sources (e.g., supermassive black 
holes, supernovae, thermal conduction, dynamical friction 
heating of orbiting satellites) using analytic and 
semi-analytic models, in addition to idealised and full 
cosmological hydrodynamic simulations (e.g., Binney \& Tabor 
1995; Narayan \& Medvedev 2001; Ruszkowski et al.\ 2004; 
McCarthy et al.\ 2004; Kim et al.\ 2005; Voit \& Donahue 
2005; Sijacki et al.\ 2008).  In most of these approaches, 
it is implicitly assumed that {\it gravitationally-induced} 
heating (e.g., via hydrodynamic shocks, turbulent mixing, 
energy-exchange between the gas and dark matter) that occurs 
during mergers/accretion is understood and treated with a 
sufficient level of accuracy.  For example, most 
current analytic and semi-analytic models of groups and 
clusters attempt 
to take the effects of gravitational heating into account by 
assuming distributions for the gas that are taken from (or 
inspired by) non-radiative cosmological hydrodynamic 
simulations --- the assumption being that these simulations 
accurately and self-consistently track gravitational processes.  

However, it has been known for some time that, even in the 
case of identical initial setups, the results of non-radiative 
cosmological simulations depend on the numerical scheme 
adopted for tracking the gas hydrodynamics.  In particular, 
mesh-based Eulerian (such as adaptive mesh refinement, 
hereafter AMR) codes appear to systematically produce higher 
entropy (lower density) gas cores within groups and clusters 
than do particle-based Lagrangian (such as smoothed particle 
hydrodynamics, hereafter SPH) codes (see, e.g., Frenk et al.\ 
1999; Ascasibar et al\. 2003; Voit, Kay, \& Bryan 2005; Dolag 
et al.\ 2005; O'Shea et al.\ 2005).  This may be regarded as 
somewhat surprising, given the ability of these codes 
to accurately reproduce a variety of different analytically 
solvable test problems (e.g., Sod shocks, Sedov blasts, the 
gravitational collapse of a uniform sphere; see, e.g., 
Tasker et al.\ 2008), although clearly hierarchical 
structure formation is a more complex and challenging test 
of the codes.  At present, the origin of the 
cores and the discrepancy in their amplitudes between 
Eulerian and Lagrangian codes in cosmologically-simulated 
groups and clusters is unclear. There have been suggestions that it
could be the result of insufficient resolution in the mesh
simulations, artificial entropy generation in the SPH simulations,
Galilean non-invariance of the mesh simulations, and differences in
the amount of mixing in the SPH and mesh simulations (e.g., Dolag et
al.\ 2005; Wadsley et al.\ 2008). We explore all of these potential
causes in \S 4.

Clearly, though, this matter is worth further investigation, 
as it potentially has important implications for the 
competition between heating and cooling in groups and 
clusters (the cooling time of the ICM has a steep dependence 
on its core entropy) and the bright end of the galaxy 
luminosity function.  And it is important to consider that 
the total heating is not merely the sum of the gravitational 
and non-gravitational heating terms.  The Rankine-Hugoniot 
jump conditions tell us that the efficiency of shock heating 
(either gravitational or non-gravitational in origin) {\it 
depends on the density of the gas at the time of heating} 
(see, e.g., the discussion in McCarthy et al.\ 2008a).  This 
implies that if gas has been heated before being shocked, 
the entropy generated in the shock can actually be amplified 
(Voit et al.\ 2003; Voit \& Ponman 2003; Borgani et al.\ 2005; 
Younger \& Bryan 2007).  The point is that gravitational and 
non-gravitational heating will couple together in complex ways, 
so it is important that we are confident that gravitational 
heating is being handled with sufficient accuracy in 
the simulations.

A major difficulty in studying the origin of the cores in 
cosmological simulations is that the group and cluster 
environment can be extremely complex, with many hundreds of 
substructures (depending on the numerical resolution)
orbiting about at any given time .  Furthermore, 
such simulations can be quite computationally-expensive if one 
wishes to resolve in detail the innermost regions of groups 
and clusters (note that, typically, the simulated cores have 
sizes $\la 0.1 r_{200}$).  An alternative approach, which we 
adopt in the present study, is to use idealised simulations of 
binary mergers to study the relevant gravitational heating 
processes.  The advantages of such an approach are obviously 
that the environment is much cleaner, therefore offering a 
better chance of isolating the key processes at 
play, and that the systems are fully resolved from the onset 
of the simulation.  The relatively modest computational 
expense of such idealised simulations also puts us in a 
position to be able to vary the relevant physical and 
numerical parameters in a systematic way and to study their 
effects.

Idealised merger simulations have been used extensively to 
study a variety of phenomena, such as the disruption of 
cooling flows (G{\'o}mez et al.\ 2002; Ritchie \& Thomas 
2002; Poole et al.\ 2008), the intrinsic scatter in cluster
X-ray and Sunyaev-Zel'dovich effect scaling relations 
(Ricker \& Sarazin 2001; Poole et al.\ 2007), the generation 
of cold fronts and related phenomena (e.g., Ascasibar \& 
Markevitch 2006), and the ram pressure stripping of orbiting 
galaxies (e.g., Mori \& Burkert 2000; McCarthy et al.\ 
2008b).  However, to our knowledge, idealised merger 
simulations have not been used to elucidate the important 
issue raised above, nor have they even been used to 
demonstrate whether or not this issue even exists in 
non-cosmological simulations.

In the present study, we perform a detailed comparison of 
idealised binary mergers run with the widely-used public 
simulations codes \flash (an AMR code) and \gadget 
(a SPH code).  The paper is organised as follows.  In \S 2 we 
give a brief description of the simulation codes and 
the relevant adopted numerical parameters.  In addition, we 
describe the initial conditions (e.g., structure of the 
merging systems, mass ratio, orbit) of our idealised 
simulations.  In \S 3, we present a detailed comparison of 
results from the \flash and \gadget runs and confirm that 
there is a significant difference in the amount of 
central entropy generated with the two codes.  In \S 4 we 
explore several possible causes for the differences we see.
Finally, in \S 5, we summarise and discuss our findings.

\section{Simulations}

\subsection{The Codes}

Below, we provide brief descriptions of the \gadget and 
\flash hydrodynamic codes used in this study and the 
parameters we have adopted.  The interested reader is referred to
Springel, Yoshida, \& White (2001) and Springel (2005b) for in-depth
descriptions of \gadget and to Fryxell et al.\ (2000) for the \flash
code. Both codes are representative examples of
their respective AMR and SPH hydrodynamic formulations, as has been 
shown in the recent code comparison of Tasker et al.\ (2008).

\subsubsection{\flash}

\flash is a publicly available AMR code developed by the Alliances
Center for Astrophysical Thermonuclear Flashes\footnote{See the flash
website at: \\ http://flash.uchicago.edu/}. Originally intended for
the study of X-ray bursts and supernovae, it has since been adapted for
many astrophysical conditions and now includes modules for relativistic
hydrodynamics, thermal conduction, radiative cooling,
magnetohydrodynamics, thermonuclear burning, self-gravity and particle
dynamics via a particle-mesh approach.  In this study we use 
\flash version 2.5.

\flash solves the Reimann problem using the piecewise 
parabolic method (PPM; Colella \& Woodward 1984).
The present work uses the default parameters, which have been 
thoroughly tested against numerous analytical tests 
(see Fryxell et al.\ 2000). The maximum number of Newton-Raphson 
iterations permitted within the Riemann solver was increased in order
to allow it to deal with particularly sharp shocks and discontinuities
whilst the default tolerance was maintained. The hydrodynamic
algorithm also adopted the default settings with periodic boundary
conditions being applied to the gas as well as to the gravity solver.

We have modified the gravity solver in \flash to use an FFT on top of
the multigrid solver (written by T. Theuns). This results in a vast
reduction in the time spent calculating the self-gravity of the
simulation relative to the publicly available version. We
have rigorously tested the new algorithm against the default multigrid
solver, more tests are presented in Tasker et al.\ (2008).

To identify regions of rapid flow change, \flash's 
refinement and de-refinement criteria can incorporate the 
adapted L\"{o}hner (1987) error estimator. This 
calculates the modified second derivative of the desired variable,
normalised by the average of its gradient over one cell.
With this applied to the density as is common place, we 
imposed the additional user-defined criteria whereby the 
density has to exceed a threshold of $200\rho_{c}$, below
which the refinement is set to the minimum $64^{3}$ mesh. 
This restricts the refinement to the interior of the clusters 
and, as we demonstrate below, was found to yield nearly 
identical results to uniform grid runs with resolution equal
to the maximum resolution in the equivalent AMR run.

\flash uses an Oct-Tree block-structured AMR grid, in which the block
to be refined is replaced by 8 blocks (in three dimensions), of which
the cell size is one half of that of the parent block. Each block
contains the same number of cells, $N_x=16$ cells in each dimension in our
runs. The maximum allowed level of refinement, $l$, is one of the
parameters of the run. At refinement level $l$, a fully refined AMR
grid will have $N_x\,2^{l-1}$ cells on a side.

All the \flash merger runs are simulated in 20 Mpc on a side periodic
boxes in a non-expanding (Newtonian) space and are run for a duration
of $\simeq 10$ Gyr. By default, our \flash simulations are run with a
maximum of $l=6$ levels of refinement ($512^3$ cells when fully
refined), corresponding to a minimum cell size of $\approx 39$ kpc,
which is small in comparison to the entropy cores produced in
non-radiative cosmological simulations of clusters (but note we
explicitly test the effects of resolution in \S 3). The simulations
include non-radiative hydrodynamics and gravity.

\subsubsection{GADGET-2}

\gadget is a publicly available TreeSPH code designed for the 
simulation of cosmological structure formation.  By default, 
the code implements the entropy 
conserving SPH formalism proposed by Springel \& Hernquist 
(2002). The code is massively parallel, has been highly 
optimised and is very memory efficient. This has led
to it being used for some of the largest cosmological 
simulations to date, including the first N-body simulation 
with more than $10^{10}$ dark matter particles (the  
{\it Millennium Simulation}; Springel et al.\ 2005a).

The SPH formalism is inherently Lagrangian in nature and 
fully adaptive, with `refinement' based on the density. The 
particles represent discrete mass elements with the fluid 
variables being obtained through a kernel interpolation 
technique (Lucy 1977; Gingold \& Monaghan 1977; Monaghan 
1992). The entropy injected through shocks is captured  
through the use of an artificial viscosity term (see, e.g., 
Monaghan 1997).  We will explore the sensitivity of our 
merger simulation results to the artificial viscosity in 
\S~\ref{viscosity}.

By default, gravity is solved through the use of a combined 
tree particle-mesh (TreePM) approach.  The TreePM method 
allows for substantial speed ups over the traditional tree 
algorithm
by calculating long range forces with the particle-mesh 
approach using Fast Fourier techniques. Higher gravitational 
spatial resolution is then achieved by applying the tree over 
small scales only, maintaining the dynamic range of the tree 
technique. This allows \gadget to vastly exceed the 
gravitational resolving power of mesh codes which rely on
the particle-mesh technique alone and are thus limited to 
the minimum cell spacing.   

We adopt the following numerical parameters by default for 
our \gadget runs (but note that most of these are 
systematically varied in \S 3).  The artificial bulk 
viscosity, $\alpha_{\rm visc}$, is set to 0.8.  The number 
of SPH smoothing neighbours, $N_{\rm sph}$, is set to 32.  
Each of our $10^{15} M_\odot$ model clusters (see \S 2.1) 
are comprised of $5\times10^5$ gas and dark matter particles 
within $r_{200}$, and the gas to total mass ratio is 
$0.141$.  Thus, the particle masses are $m_{\rm gas} = 2.83 
\times 10^8 M_\odot$ and $m_{\rm dm} = 1.72 \times 10^{9} 
M_\odot$.  The gravitational softening length is set to 10 
kpc, which corresponds to $\approx 5 \times 10^{-3} r_{200}$ 
initially.  All the \gadget merger runs are simulated in 20 
Mpc on a side periodic boxes in a non-expanding (Newtonian) 
space and are run for a duration of $\simeq$ 10 Gyr.  The 
simulations include basic hydrodynamics only (i.e., are 
non-radiative).

\subsection{Initial Conditions}

In our simulations, the galaxy clusters are initially 
represented by spherically-symmetric systems composed of a 
realistic mixture of dark matter and gas.

The dark matter is assumed to follow a NFW distribution
(Navarro et al.\ 1996; 1997):

\begin{equation}
\rho(r) = \frac{\rho_s}{(r/r_s)(1+r/r_s)^2}
\end{equation}

\noindent where $\rho_s = M_s/(4 \pi r_s^3)$ and

\begin{equation}
M_s=\frac{M_{200}}{\ln(1+r_{200}/r_s)-(r_{200}/r_s)/(1+r_{200}/r_s)} \ \ .
\end{equation}

Here, $r_{200}$ is the radius within which the mean
density is 200 times the critical density, $\rho_{\rm crit}$, 
and $M_{200} \equiv M(r_{200}) = (4/3) \pi r_{200}^3 \times 
200 \rho_{\rm crit}$.

The dark matter distribution is fully specified once an 
appropriate scale radius ($r_s$) is selected.  The scale 
radius can be expressed in terms of the halo concentration 
$c_{200} = r_{200}/r_s$.  We adopt a concentration of 
$c_{200} = 4$ for all our systems.  This value is typical of 
massive clusters formed in $\Lambda$CDM cosmological 
simulations (e.g., Neto et al.\ 2007).

In order to maintain the desired NFW configuration, 
appropriate velocities must be assigned to each dark matter 
particle.  For this, we follow the method outlined in McCarthy 
et al.\ (2007).  Briefly, the three velocity components are 
selected randomly from a Gaussian distribution whose width is 
given by the local velocity dispersion [i.e, $\sigma(r)$].  
The velocity dispersion profile itself is determined by 
solving the Jeans equation for the mass density distribution 
given in eqn.\ (1).  As in McCarthy et al.\ (2007), the dark 
matter haloes are run separately in isolation in \gadget
for many dynamical times to ensure that they have fully relaxed.

For the gaseous component, we assume a powerlaw 
configuration for the entropy\footnote{Note, the quantity $K$ 
is the not the actual thermodynamic specific entropy ($s$) of 
the gas, but is related to it via the simple relation $s 
\propto \ln{K^{3/2}}$.  However, for historical reasons we 
will refer to $K$ as the entropy.}, $K \equiv P \rho_{\rm 
gas}^{-5/3}$, by default.  In particular,

\begin{equation}
\frac{K(r)}{K_{200}} = 1.47 \biggl(\frac{r}{r_{200}} \biggr)^{1.22}\,,
\end{equation}

\noindent where the `virial entropy', $K_{200}$, is given by
\begin{equation}
K_{200} \equiv \frac{G M_{200}}{2 \ r_{200}} \frac{1}{(200 
\ \rho_{\rm crit})^{2/3}}\,.
\end{equation}

This distribution matches the entropy profiles of groups and 
clusters formed in the non-radiative cosmological simulations 
of Voit, Kay, \& Bryan (2005) (VKB05) for $r \ga 0.1 r_{200}$.  It 
is noteworthy that VKB05 find that 
this 
distribution approximately matches the entropy profiles of 
both SPH (the \gadget code) and AMR (the \enzo code) 
simulations.  Within $0.1 r_{200}$, however, the AMR and SPH 
simulations show evidence for entropy cores, but of 
systematically different amplitudes.  We initialise our 
systems without an entropy core [i.e., eqn. 
(3) is assumed to hold over all radii initially] to see, 
first, if such cores are established during the merging 
process and, if so, whether the amplitudes differ between the 
SPH and AMR runs, as they do in cosmological simulations.  We 
leave it for future work to explore the differences that 
result (if any) between SPH and AMR codes when large cores are 
already present in the initial systems (e.g., the merger of 
two `non-cool core' systems).

With the mass distribution of dark matter established (i.e., 
after having run the dark matter haloes in isolation) and an 
entropy distribution for the gas given by eqn.\ (3), we 
numerically compute the radial gas pressure profile (and 
therefore also the gas density and temperature profiles), 
taking into account the self-gravity of the gas, by 
simultaneously solving the equations of hydrostatic 
equilibrium and mass continuity:

\begin{eqnarray}
\frac{{\rm dlog}P}{{\rm dlog}M_{\rm gas}}= - \frac{G M_{\rm
gas} M_{\rm tot}}{4 \pi r^4 P} \\
\frac{{\rm dlog}r}{{\rm dlog}M_{\rm gas}}= \frac{M_{\rm gas}}{4 
\pi r^3} \biggl(\frac{K}{P} \biggr)^{3/5}
\end{eqnarray} 

Two boundary conditions are required to solve these equations. 
The first condition is that $r(M_{\rm gas}=0) = 0$.   
The second condition is that the total mass of hot gas 
within $r_{200}$ yields a realistic baryon fraction of 
$M_{\rm gas}/M_{tot} = 0.141$. In order to meet the second
condition, we choose a value for $P(M_{\rm gas}=0)$ and
propagate the solution outwards to $r_{200}$. We then 
iteratively vary the inner pressure until the desired baryon 
fraction is achieved.

For the \gadget simulations, the gas particle positions 
are assigned by radially morphing a glass distribution until 
the desired gas mass profile is obtained (see McCarthy et 
al.\ 2007).  The entropy (or equivalently internal energy per 
unit mass) of each particle is specified by interpolating 
with eqn.\ (3).  For the \flash simulations, the gas 
density and entropy of each grid cell is computed by using 
eqn.\ (3) and interpolating within the radial gas density 
profile resulting from the solution of eqns.\ (5) and (6).

Thus, for both the \gadget and \flash simulations 
we start with identical dark matter haloes (using particle 
positions and velocities from the \gadget isolated runs  
with a gravitational softening length of 10kpc) and 
gas haloes, which have been established by interpolating 
within radial profiles that have been computed numerically 
under the assumption that the gas is initially in hydrostatic 
equilibrium within the dark matter halo. Note that when 
varying the resolution of the \flash simulations we simply change 
the maximum number of refinements $l$ --- we do not vary the number 
of dark matter particles --- in this way, the initial dark matter 
distribution is always the same as in the low resolution \gadget run
in all the simulations we have run.

In both the \gadget and \flash simulations, the gaseous haloes 
are surrounded by 
a low density pressure-confining gaseous medium that prevents 
the systems from expanding prior to the collision (i.e., so 
that in the case of an isolated halo the object would be 
static) but otherwise it has a negligible dynamical effect on 
the system.
 
Isolated gas+DM haloes were run in both \gadget and \flash for
10 Gyr in order to test the stability of the initial gas and dark
matter haloes.  Although deviations in the central entropy develop over 
the course of the isolated simulations, indicating the systems are not 
in perfect equilibrium initially, they are small in amplitude (the 
central entropy increases by $<10\%$ over 10 Gyr), especially in 
comparison to the factor of $\sim2-3$ jump in the central entropy that 
occurs as a result of shock heating during the merger.  Furthermore, we 
note that the amplitude of the deviations in the isolated runs are 
significantly decreased as the resolution of these runs is increased.  
Our merger simulations, however, are numerically converged (see \S 3), 
indicating that the deviations have a negligible effect on merger 
simulation results and the conclusions we have drawn from them.

%--------------------------------------------------------------------------

\begin{figure*}
\centering
\leavevmode
\epsfysize=8.4cm \epsfbox{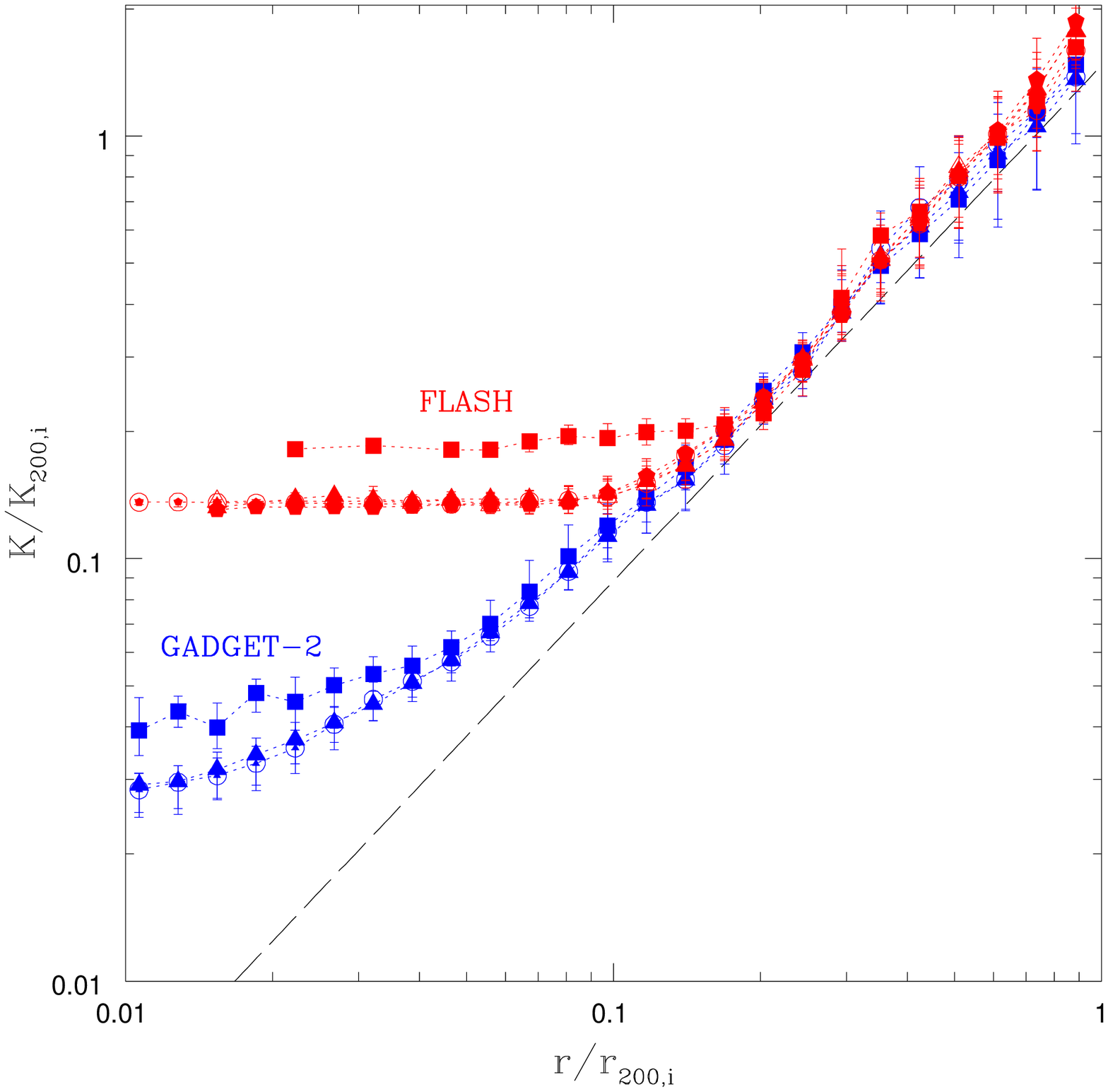}
\epsfysize=8.4cm \epsfbox{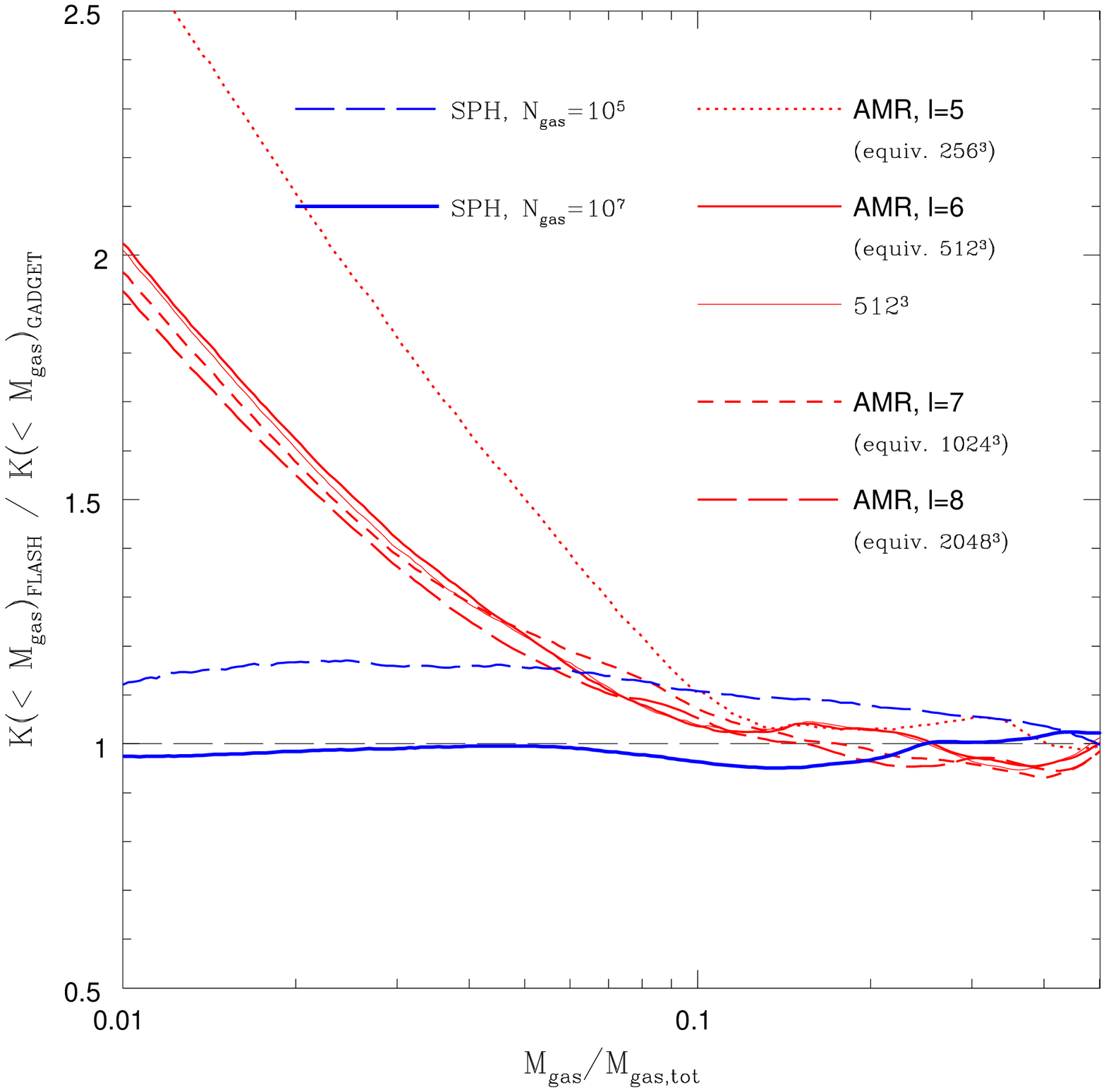}
\caption{
Plots demonstrating the entropy cores formed in idealised 
head on mergers of equal mass ($10^{15} M_\odot$) clusters 
in the \flash and \gadget simulations.  The left hand panel 
shows the final radial entropy distribution, 
where the data points are the median entropy value in radial 
bins and the error bars correspond to the 25th and 75th 
percentiles.  The dashed black line represents the initial 
powerlaw configuration.  The solid blue squares, solid 
blue triangles, and open circles represent the low resolution 
($10^{5}$ gas particles), default ($10^{6}$ gas particles), and 
high resolution ($10^{7}$ gas particles) \gadget runs.  The 
minimum SPH smoothing lengths of these simulations throughout the 
runs are approximately $25$, $11$, and $5$ kpc, respectively.  
The solid red squares, solid red triangles, 
solid red pentagons, and open red circles represent \flash 
AMR runs with $l =$ 5, 6 (default), 7, and 8, 
respectively, which have minimum cell sizes of $\approx 78$, 
$39$, $19.5$, and $9.8$ kpc (respectively).  (These would be
equivalent to the resolutions of uniform grid runs with 
$256^3$, $512^3$, $1024^3$ and $2048^3$ cells.)  The open red 
triangles represent a uniform $512^3$ \flash run with a cell 
size of $\approx 39$ kpc (for reference, $r_{200,i} \simeq 
2062$ kpc).  With the exception 
of the lowest resolution AMR run, all of the \flash runs 
essentially lie on top of one another, as do the Gadget 
runs, meaning
both runs are numerically converged. However, importantly 
the two codes have converged to results that differ by a 
factor $\sim 2$ in central entropy.
The right hand panel presents the results in a slightly 
different way: it shows the entropy as a function of
enclosed gas mass $K(<M_{\rm gas})$.  This is constructed by
simply sorting the particles/cells by entropy in ascending
order and then summing masses of the particles/cells.  The 
results have been 
normalised to the final distribution of the default \gadget 
run (dashed black line).  The dashed blue and solid blue curves 
represent the low and high resolution \gadget runs, respectively, whereas
the dotted red, 
solid red, short-dashed red, and long-dashed red curves  
represent the \flash AMR runs $l =$ 5, 6, 7, 
and 8.  The thin solid red curve represents the uniform 
$512^3$ \flash run.  Again we see that the default \gadget 
and \flash runs are effectively converged, but to a 
significantly different profile.} 
\label{entropy_radius}
\end{figure*}

\section{Idealised cluster mergers}
\label{thecomparison}

The existence of a discrepancy between the inner properties 
of the gas in groups and clusters formed in AMR and SPH 
cosmological simulations was first noticed in the
Santa Barbara code comparison of Frenk et al.\ (1999).  It was 
subsequently verified in several works, including Dolag et 
al.\ (2005), O'Shea et al.\ (2005), Kravtsov, Nagai \& 
Vikhlinin (2005), and VKB05.  The latter study 
in particular clearly demonstrated, using a relatively large 
sample of 
$\sim 60$ simulated groups and clusters, that those systems 
formed in the AMR simulations had systematically larger 
entropy cores than their SPH counterparts.  Since 
this effect was observed in cosmological simulations, it was 
generally thought that the discrepancy was due to 
insufficient resolution in the mesh codes at high 
redshift (we note, however, that VKB05 argued against resolution 
being the cause).  This would result in under-resolved small scale 
structure formation in the early universe.  This explanation 
is consistent with the fact that in the Santa 
Barbara comparison the entropy core amplitude tended to be 
larger for the lower resolution mesh code runs.  Our first aim is 
therefore to determine 
whether the effect is indeed due to resolution limitations, 
or if it is due to a more fundamental difference between the 
two types of code.  We test this using identical idealised 
binary mergers of spherically-symmetric clusters in \gadget 
and \flash, where it is possible to explore the effects of 
finite resolution with relatively modest computational 
expense (compared to full cosmological simulations).

\subsection{A Significant Discrepancy}
\label{asignificantdescrepancy}

As a starting point, we investigate the generation of entropy 
cores in a head on merger between two identical $10^{15} 
M_\odot$ clusters, each colliding with an initial speed of 
$0.5 V_{\rm circ}(r_{200}) \simeq 722$ km/s [i.e., the 
initial 
{\it relative} velocity is $V_{\rm circ}(r_{200})$, which is 
typical of merging systems in cosmological simulations; 
see, e.g., Benson 2005]. The system is initialised such that 
the two clusters are just barely touching (i.e., their 
centres are separated by $2 r_{200}$).  The simulations are 
run for a duration of 10 Gyr, by the end of which the merged 
system has relaxed and there is very little entropy generation 
ongoing.

\begin{table}
\caption{Characteristics of the head on simulations presented in
\S 3.1}
\centering
\begin{tabular}{ l l l}
\hline
FLASH sim. & No. cells & Max. spatial res.\\
 & & (kpc)\\
\hline
$l=5$ & equiv.\ $256^3$ & 78 \\
$l=6$ (default) &  equiv.\ $512^3$ & 39 \\
$l=7$ &  equiv.\ $1024^3$ & 19.5 \\
$l=8$ &  equiv.\ $2048^3$ & 9.8 \\
$512^3$ &  $512^3$ & 39 \\
\hline
\\
\hline
GADGET-2 sim. & No. gas particles & Max. spatial res.\\
 & & (kpc)\\
\hline
low res. & $10^5$ & $\approx 25$ \\
default & $10^6$ & $\approx 11$ \\
hi res. & $10^7$ & $\approx 5$ \\
\hline
\end{tabular} 
\end{table}

Our idealised test gives very similar results to non-radiative 
cosmological simulations --- there is a distinct 
difference in the amplitude of the entropy cores in the AMR 
and SPH simulations, with the entropy in the mesh code a 
factor $\sim 2$ higher than the SPH code. It is evident that
the difference between the codes is captured in a single merger
event. An immediate 
question is whether this is the result of the different
effective resolutions of the codes.  Resolution tests can be 
seen in the left hand panel of Figure~\ref{entropy_radius}, 
where we plot the resulting radial entropy distributions. 
For \gadget, we compare runs with $10^5$, $10^6$ (the default), 
and $10^7$ particles. For \flash we compare AMR runs with 
minimum cell sizes of $\approx 78$, $39$ (the default), $19.5$, and 
$9.8$ kpc and a uniform grid run with the default 39~kpc cell size.
The simulation characteristics for these head on mergers are 
presented in Table 1.
To make a direct comparison with the cosmological results of 
VKB05 (see their Fig.\ 5), we normalise the entropy by the 
initial `virial' entropy ($K_{200}$; see eqn.\ 4) and the 
radius by the initial virial radius, $r_{200}$.

The plot clearly shows that the simulations converge on two 
distinctly different solutions within the inner ten percent 
of $r_{200}$, whereas the entropy at large radii shows 
relatively good agreement between the two codes.  The 
simulations performed for the resolution test span a factor 
of 8 in spatial resolution in \flash and approximately a factor of 
5 in \gadget.  The \flash AMR runs effectively 
converge after reaching a peak resolution equivalent to a  
$512^{3}$ run (i.e., a peak spatial resolution of $\approx 
39$ kpc or $\approx 0.019 r_{200}$).  We have also tried a \flash run with 
a uniform (as opposed to adaptive) $512^3$ grid and the results 
essentially trace the AMR run with an equivalent peak resolution.  This 
reassures us that our AMR refinement criteria is correctly capturing
all regions of significance.  The lowest resolution SPH run, which only 
has $5\times10^4$ gas particles within $r_{200}$ initially, has a slightly 
higher final central entropy than the default and high resolution SPH 
runs.  This may not be surprising given the tests and modelling presented
in Steinmetz \& White (1997).  These authors demonstrated that with such 
small particle numbers, two-body heating will be important if the mass of 
a dark matter particle is significantly above the mass of a gas particle. 
The \gadget runs converge, however, when the number of gas and dark 
matter particles are increased by an order magnitude (i.e., as in our 
default run), yielding a maximum spatial resolution of $\approx 11$ kpc 
(here we use the minimum SPH smoothing length as a measure of the maximum 
spatial resolution).

\begin{figure*}
\centering
\includegraphics[width=13.5cm]{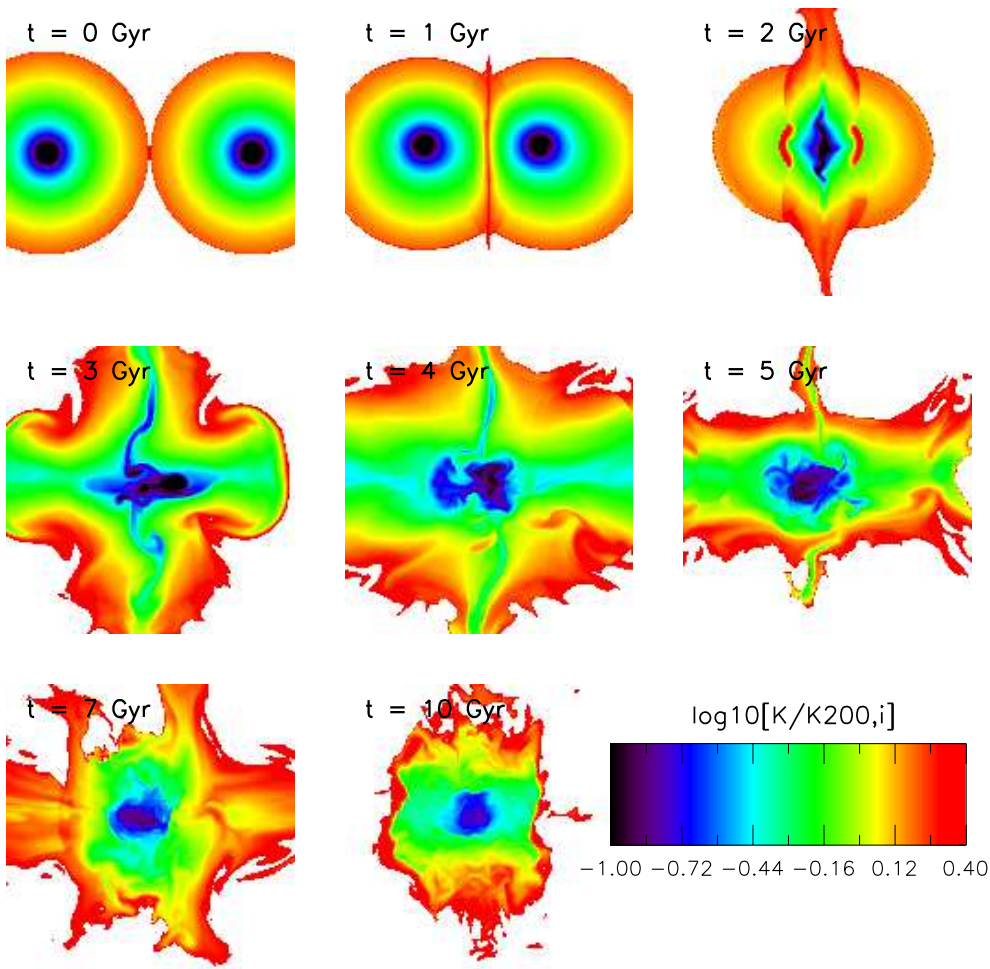}
\caption{Logarithmic entropy slices (i.e., thickness of 
zero) through the centre of
the default \flash merger simulation (with $l=6$) at times 0, 1, 2, 3, 4,
5, 7 and 10 Gyr. The lowest
entropy material is shown in blue, increasing in entropy
through green, yellow to red.  Each panel is 6 Mpc on a side.
Significant entropy is generated at $t\approx 2$~Gyr when 
the cores collide and gas is squirted out, and again later on 
when this gas reaccretes. 
 }
\label{8panelimage}
\end{figure*}

A comparison of the left hand panel of Fig.\ 1 to Fig.\ 5 of VKB05 
reveals a remarkable 
correspondence between the results of our idealised merger 
simulations and those of their cosmological simulations 
(which spanned system masses of $\sim 10^{13-15} M_\odot$).  
They find that the ratio of the AMR and SPH core amplitudes is 
$\sim 2$ in both the idealised and cosmological 
simulations. This difference is also seen in the Santa Barbara 
comparison of Frenk et al.\ (1999) when comparisons are made 
between the SPH simulations and the highest resolution AMR 
simulations carried out in that study (ie., the `Bryan' AMR 
results)\footnote{We note, however, that the lower resolution AMR 
simulations in that study produced larger entropy cores, which 
suggests that they may not have been numerically converged (as in 
the case of $l=5$ AMR run in Fig.\ 1).}. This consistency 
presumably indicates that whatever mechanism is 
responsible for the differing core amplitudes in 
the cosmological simulations is also responsible for the 
differing core amplitudes in our idealised simulations.  
This is encouraging, as it implies the generation of the 
entropy cores can be studied with idealised simulations.  
As outlined in \S 1, the advantage of idealised 
simulations over cosmological simulations is their relative 
simplicity.  This gives us hope that we can use idealised 
simulations to track down the underlying cause of the 
discrepancy between particle-based and mesh-based 
hydrodynamic codes.

The right hand panel of Figure~\ref{entropy_radius} shows 
the resulting entropy distributions plotted in a slightly 
different fashion.  Here we plot the entropy as a function of 
`enclosed' gas mass $K(<M_{\rm gas})$.  This is constructed 
by simply ranking the particles/cells by entropy in ascending 
order and then summing the masses of the particles/cells [the 
inverse, $M_{\rm gas}(K)$, would therefore be the total mass 
of gas with entropy lower than $K$].  Convective stability 
ensures that, eventually when the system is fully relaxed, 
the lowest-entropy gas will be located at the very centre of 
the potential well, while the highest entropy gas will be 
located at the system periphery.  $K(<M_{\rm gas})$ is 
therefore arguably a more fundamental quantity than $K(r)$ 
and we adopt this test throughout the rest of the paper.  It 
is also noteworthy that in order to compute $K(<M_{\rm 
gas})$ one does not first need to select a reference point 
(e.g., the centre of mass or the position of the particle 
with the lowest potential energy) or to bin the 
particles/cells in any way, both of which could 
introduce ambiguities in the comparison between the SPH and 
AMR simulations (albeit likely minor ones).

In the right hand panel of Figure~\ref{entropy_radius}, we 
plot the resulting $K(<M_{\rm gas})$ distributions normalised 
to the final entropy distribution of the default \gadget 
run.  Here we see that the lowest-entropy gas in the 
\flash runs have a higher entropy, by a factor of $\approx 
1.9-2.0$, than the lowest-entropy gas in the default \gadget run.  
Naively, looking at the right hand panel of 
Figure~\ref{entropy_radius} one might conclude that the 
discrepancy is fairly minor, given that $\sim 95$\% of the 
gas has been heated to a similar degree in the SPH and AMR 
simulations.  But it is important to keep in mind that it is 
the properties of the lowest-entropy gas in particular that are 
most relevant to the issue of heating vs.\ cooling in groups 
and clusters (and indeed in haloes of all masses), since 
this is the gas that has the shortest cooling time.

The agreement between our results and those from  cosmological 
simulations (e.g., Frenk et al.\ 1999; VKB05) is striking.  The
convergence of the entropy distributions in our idealised simulations
negates the explanation that inadequate resolution of the high
redshift universe in cosmological AMR simulations is the root cause of
the discrepancy between the entropy cores in SPH and AMR 
simulations (although we note that some of the lower 
resolution AMR simulations in the study of Frenk et al.\ may not 
have been fully converged and therefore the discrepancy may 
have been somewhat exaggerated in that study for those 
simulations).
We therefore conclude that the higher entropy generation in AMR codes
relative to SPH codes within the cores of groups and clusters arises
out of a more fundamental difference in the adopted algorithms.  Below
we examine in more detail how the entropy is generated during the
merging process in the simulations and we then systematically explore
several possible causes for the differences in the simulations.

\subsection{An overview of heating in the simulations}
\label{overview}

We have demonstrated that the entropy generation that takes 
place in our idealised mergers is robust to our choice of 
resolution, yet a difference persists in the amount of 
central entropy that is generated in the SPH and mesh
simulations.  We now examine the entropy generation as a 
function of time in the simulations, which may provide clues 
to the origin of the difference between the codes.

\begin{figure*}
\centering
\leavevmode
\epsfysize=6.0cm \epsfbox{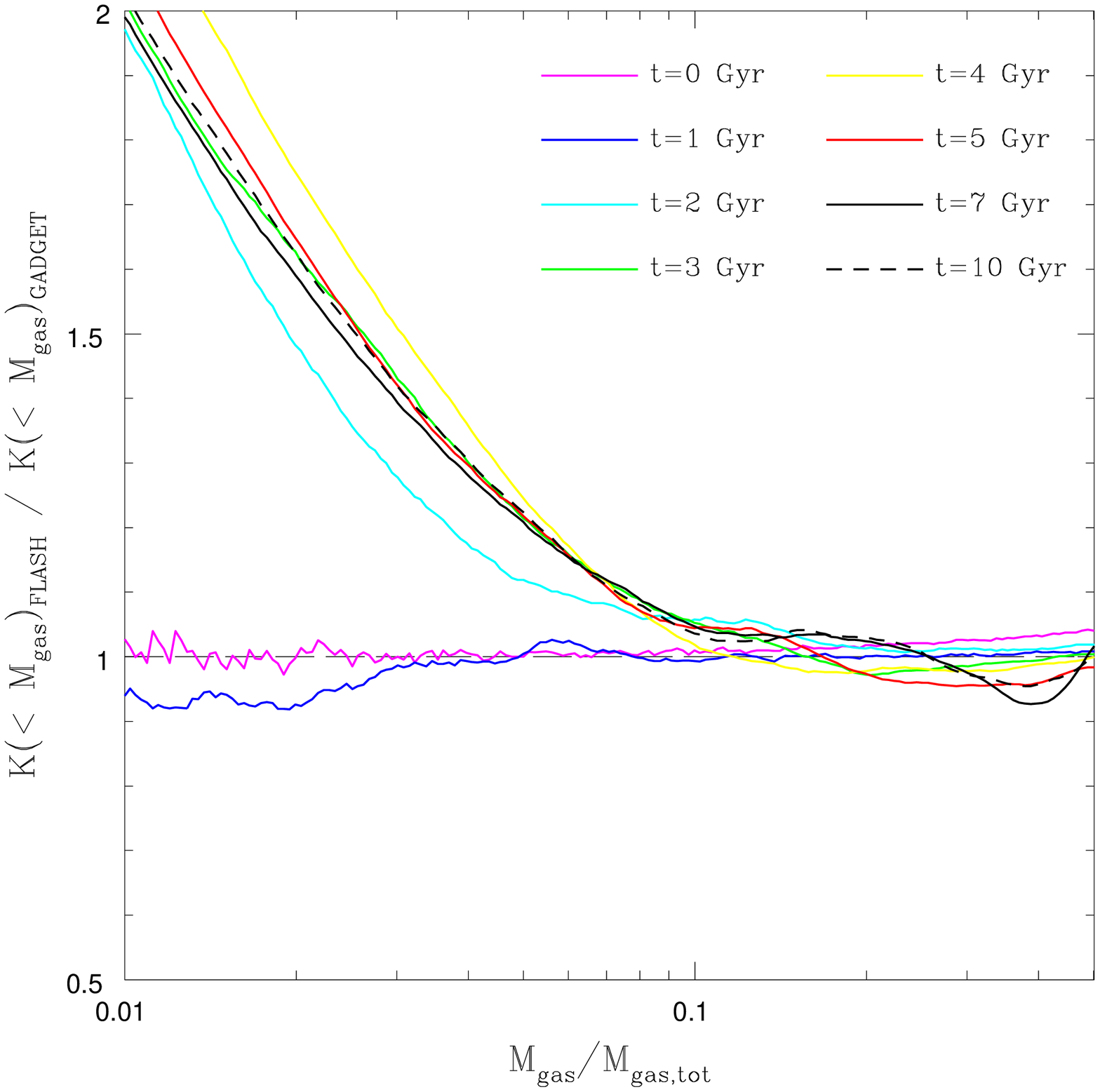}
\epsfysize=6.0cm \epsfbox{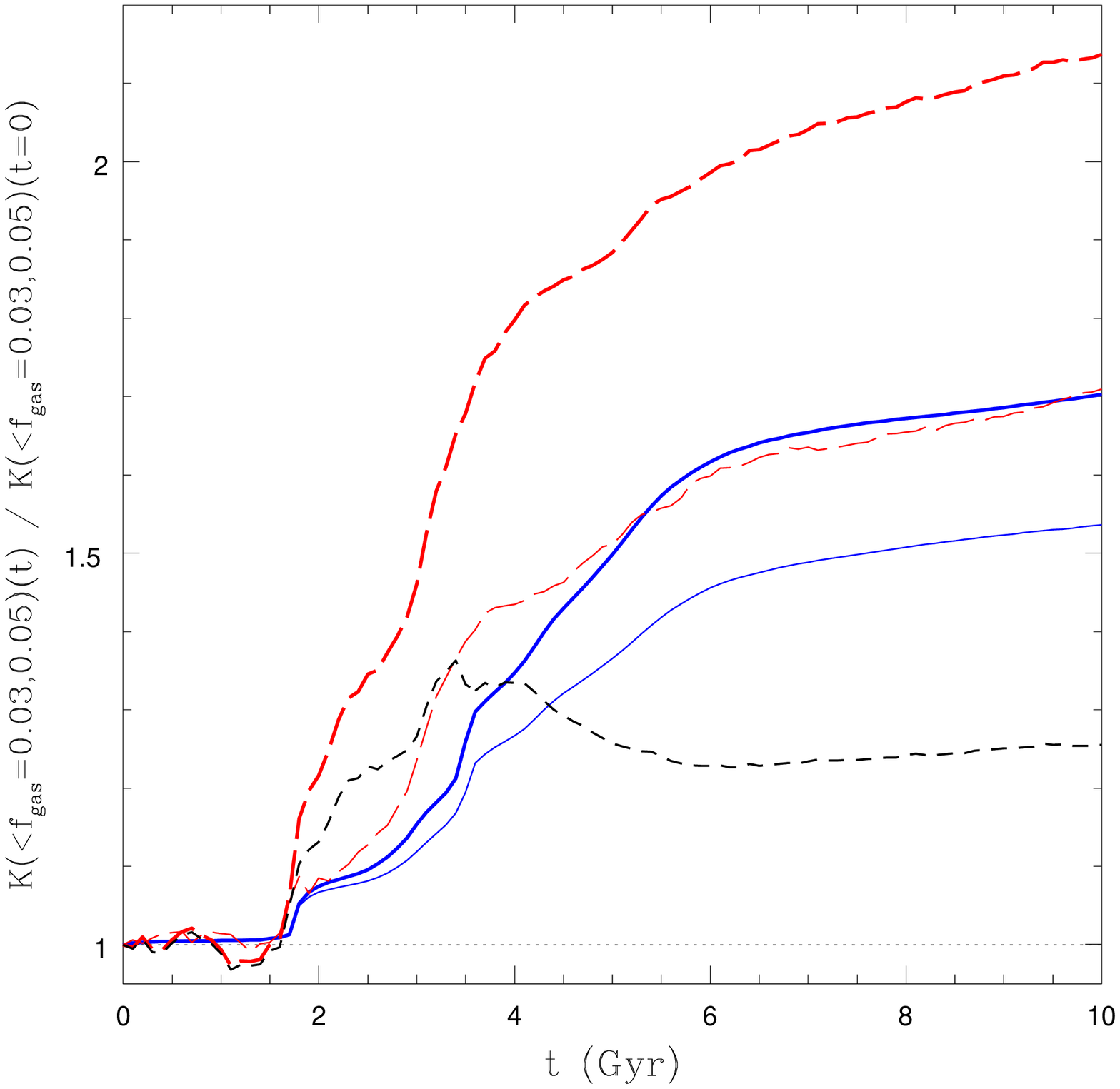}
\epsfysize=6.0cm \epsfbox{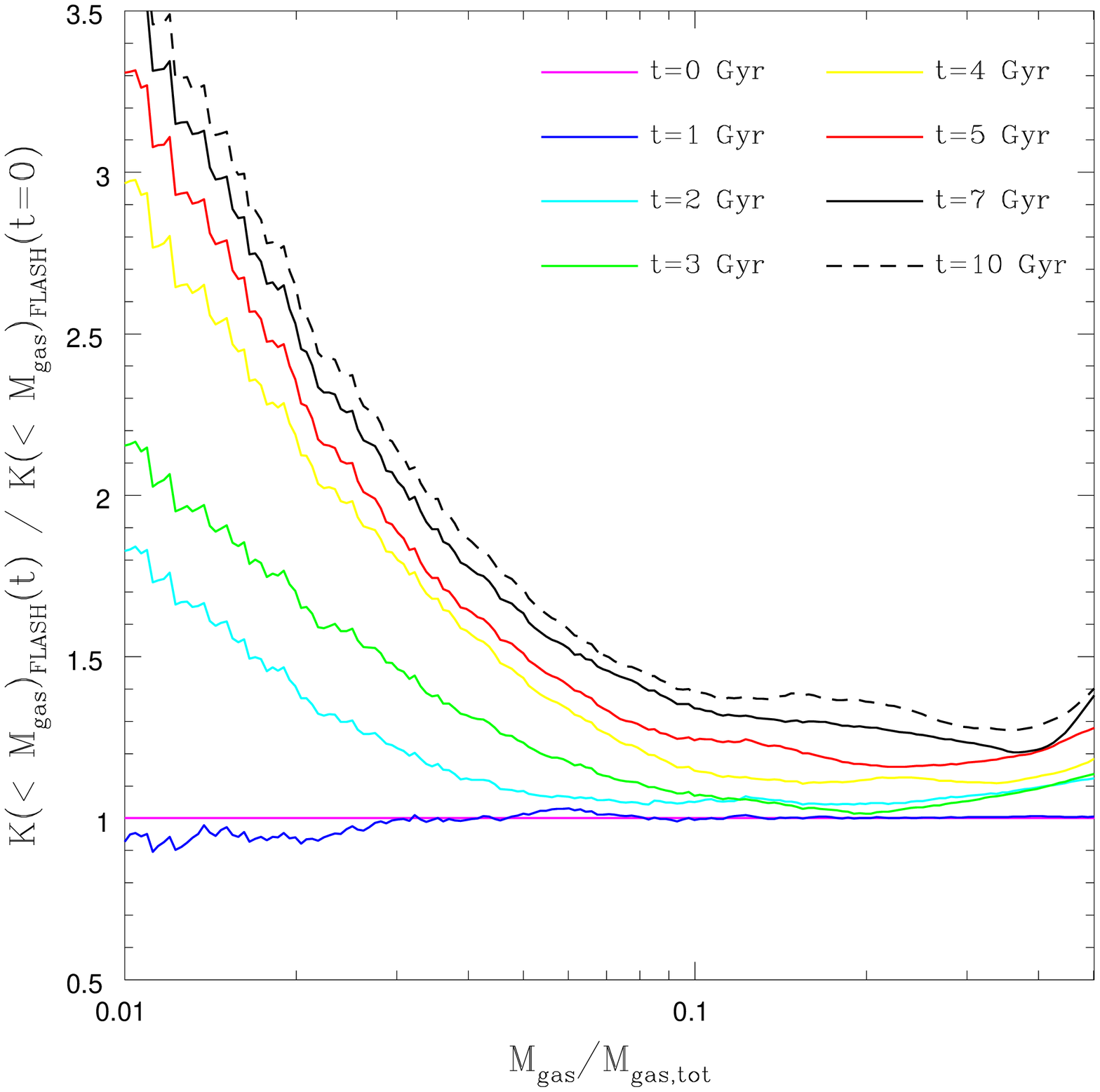}
\epsfysize=6.0cm \epsfbox{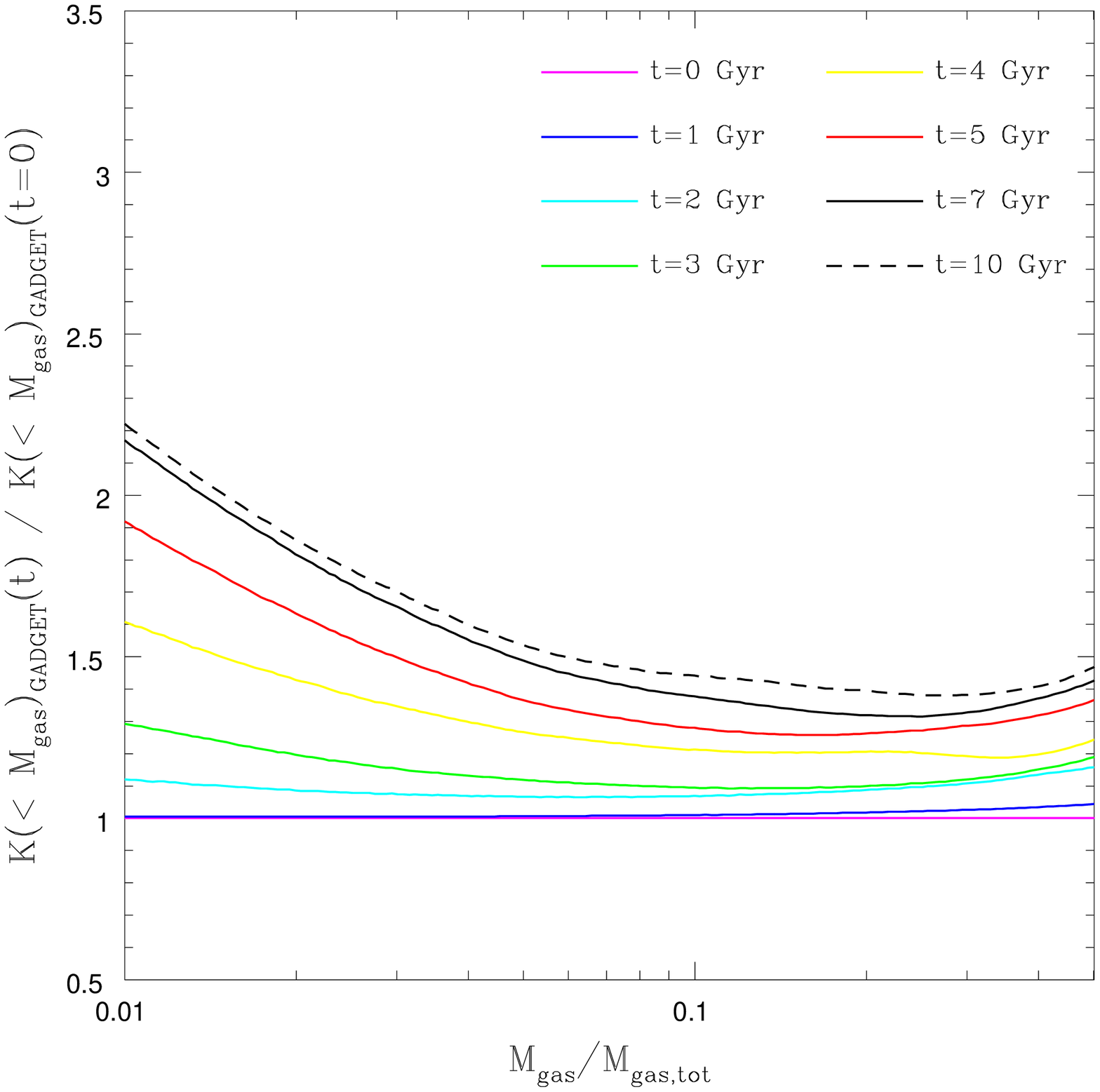}
\caption{The time-dependence of entropy generation in the
default \gadget and \flash runs.  The top left panel shows
the ratio of $K(<M_{\rm gas})$ in the default \flash run to
$K(<M_{\rm gas})$ in the default \gadget run.  The various
curves represent the ratio at different times during the
simulations (see legend).  The top right panel shows the time
evolution of the lowest-entropy gas only in the default runs.
Shown are $K(<M_{\rm gas}/M_{\rm gas,tot} = 0.03)$ (thick
curves) and $K(<M_{\rm gas}/M_{\rm gas,tot} = 0.05)$ (thin
curves) for the \flash (long-dashed red curves) and \gadget
(solid blue curves) runs (i.e., having sorted the gas
particles/cells by entropy, we show the evolution of the
entropy that encloses 3\% and 5\% of the total gas mass).
The curves have been normalised to their initial values at
the start of the simulations.  The short-dashed black curve
represents the ratio of \flash to \gadget entropies enclosing
3\% of the total gas mass.  The bottom two panels show the 
$K(<M_{\rm gas})$ distributions for the default \flash and 
\gadget runs separately, at different times during the 
simulation.  Together, these plots illustrate that the 
difference in the final entropy distributions of the \flash 
and \gadget runs is primarily established around the time of 
core collision ($\sim 2-3$ Gyr).  It is worth noting, however, 
that significant entropy generation continues after this 
time, but it occurs in nearly the same fashion in 
the AMR and SPH runs. }
\label{entropygastime}
\end{figure*}

Figure~\ref{8panelimage} shows $\log(K)$ in a slice
through the centre of the default \flash simulation at times 
0, 1, 2, 3, 4, 5, 7 and 10 Gyr. This may be compared to 
Figure~\ref{entropygastime}, which shows the entropy 
distribution of the simulations as a function of time (this 
figure is described in detail below).  Briefly, as the cores 
approach each other, a relatively gentle shock front forms 
between the touching edges of the clusters, with gas being 
forced out 
perpendicular to the collision axis.  Strong heating does 
not actually occur until approximately the time when the 
cores collide, roughly 1.8 Gyrs into the run.  The shock 
generated through the core collision propagates outwards, 
heating material in the outer regions of the system.  This 
heating causes the gas to expand and actually overshoot 
hydrostatic equilibrium.  Eventually, the gas, which 
remains gravitationally bound to the system, begins to fall 
back onto the system, producing a series of weaker secondary 
shocks.  Gas at the outskirts of the system, which is the 
least bound, takes the longest to re-accrete.  This 
dependence of the time for gas to be re-accreted upon the 
distance from the centre results in a more gradual increase 
in entropy than seen in the initial core collision. 
In a {\it qualitative} sense, the heating process that takes 
place in the \flash simulations is therefore very similar to 
that seen in the \gadget simulations (see \S 3 of McCarthy et 
al.\ 2007 for an overview of the entropy evolution in 
idealised \gadget mergers). 

The top left panel in Figure~\ref{entropygastime} shows the 
ratio of $K(<M_{\rm gas})$ in the default \flash run relative 
to $K(<M_{\rm gas})$ in the default \gadget run.  The various 
curves represent the ratio at different times during the 
simulations (see figure key --- note that these correspond 
to the same outputs displayed in Figure~\ref{8panelimage}).  
It can clearly be seen that the bulk of the difference in 
the {\it final} entropy distributions of the simulations is 
established around the time of core collision.  The ratio of 
the central entropy in the \flash simulation to the central 
entropy in the \gadget simulation converges after $\approx 
4$ Gyr.  The top right panel shows the time evolution of the 
lowest-entropy gas only in both the \gadget and \flash runs.  
Here we see there are similar trends with time, in the sense 
that there are two main entropy generation episodes (core 
collision and re-accretion), but that the entropy 
generated in the first event is much larger in the \flash run 
than in the \gadget run.  Far outside the core, however, the 
results are very similar.  For completeness, the bottom two 
panels show $K(<M_{\rm gas})$ at different times for the 
\gadget and \flash runs separately. 

The small initial drop in the central entropy at 1 Gyr in the 
\flash run (see bottom left panel) is most likely due to 
interpolation errors at low resolution.  This drop in entropy 
should not physically occur without cooling processes (which are 
not included in our simulations), but there is nothing to prevent a 
dip from occurring in the simulations due to numerical inaccuracies 
(the second law of thermodynamics is not explicitly hardwired into 
the mesh code).  At low resolutions, small violations in entropy 
conservation can occur due to inaccurate interpolations made by the 
code.  We have verified that the small drop in entropy does not 
occur in the higher resolution \flash runs.  We note that while 
this effect is present in default \flash run, it is small and as 
demonstrated in Fig.\ 1 the default run is numerically converged. 

It is interesting that the \flash to \gadget central entropy 
ratio converges relatively early on in the simulations.  This 
is in spite of the fact that a significant fraction of the 
entropy that is generated in both simulations is actually 
generated at later times, during the re-accretion phase.  
Evidently, this phase occurs in a very similar fashion in 
both simulations.  In \S 4, we will return to the point that 
the difference between the results of the AMR and SPH 
simulations arises around the time of core collision.

\subsection{Alternative setups}

It is important to verify that the conclusions we have drawn 
from our default setup are not unique to that specific 
initial configuration.  Using a suite of merger simulations 
of varying mass ratio and orbital parameters, McCarthy et 
al.\ (2007) demonstrated that the entropy generation that 
takes place does so in a qualitatively similar manner to 
that described above in all their simulations.  However, 
these authors examined only SPH simulations.  We have 
therefore run several additional merger simulations in both 
\gadget and \flash to check the robustness of our conclusions.
All of these mergers are carried out using the same resolution
as adopted for the default \gadget and \flash runs.

\begin{figure}
\centering
\includegraphics[width=8.4cm]{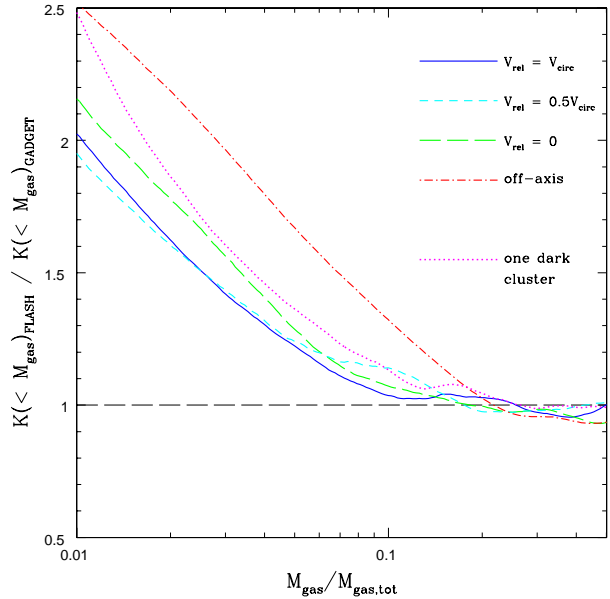}
\caption{The ratio of \flash to \gadget final entropy
distributions for equal mass mergers of varying initial 
orbital parameters.  The solid blue curve represents the 
default setup (head on collision with an initial relative 
velocity of $V_{\rm circ}(r_{200})$).  The long-dashed green 
and short-dashed cyan curves represent head on collisions 
with initial relative velocities of $0.5
V_{\rm circ}(r_{200})$ and $0$ (i.e., at rest initially).
The dot dashed red curve represents an off-axis collision
with an initial relative radial velocity of $\simeq 0.95
V_{\rm circ}(r_{200})$ and an initial relative tangential
velocity of $\simeq 0.312 V_{\rm circ}(r_{200})$ (i.e., the
total energy is equivalent to that of the default setup).
Also shown (dotted magenta curve), is
the entropy ratio of a run where one of the clusters is
composed of dark matter only and the other of a realistic
mixture of gas and dark matter (see text).  All these 
simulations result in a comparable difference in entropy 
profile between the mesh code and the SPH code.
 } \label{orbit_test}
\end{figure}

In Figure~\ref{orbit_test}, we plot the final \flash to 
\gadget $K(<M_{\rm gas})$ ratio for equal mass mergers with 
varying orbital parameters (see figure caption).  In all 
cases, \flash systematically produces larger entropy cores 
than \gadget, and by a similar factor to that seen in the 
default merger setup.  Interestingly, the off-axis case 
results in a somewhat larger central entropy discrepancy 
between \gadget and \flash, even though the bulk energetics 
of this merger are the same as for the default case.  A 
fundamental difference between the off-axis case and the 
default run is that the former takes a longer time for the 
cores to collide and subsequently relax (but note 
by the end of the off-axis simulation there is very little 
ongoing entropy generation, as in the default case).  This 
may suggest that the timescale over which entropy is 
generated plays some role in setting the magnitude of the 
discrepancy between the AMR and SPH simulations.  For 
example, one possibility is that `pre-shocking' due to the 
artificial viscosity i.e., entropy generation during the early phases 
of the collision when the interaction is subsonic or mildly transonic)
in the SPH simulations becomes more 
relevant over longer timescales.  Another possibility is that 
mixing, which is expected to be more prevalent in Eulerian 
mesh simulations than in SPH simulations, plays a larger 
role if the two clusters spend more time in orbit about each 
other before relaxing into a single merged system (of 
course, one also expects enhanced mixing in the off-axis 
case simply because of the geometry).  We explore these and 
other possible causes of the difference in \S 4.

In addition to varying the orbital parameters, we have also 
experimented with colliding a cluster composed of dark matter 
only with another cluster composed of a realistic 
mixture of gas and dark matter (in this case, we 
simulated the head on merger of two equal mass $10^{15} 
M_\odot$ clusters with an initial relative velocity of 
$\simeq 1444$ km/s). Obviously, this is not an 
astrophysically reasonable setup.  However, a number of 
studies have suggested that there is a link between the 
entropy core in clusters formed in non-radiative 
cosmological simulations and the amount of 
energy exchanged between the gas and the dark matter in 
these systems (e.g., Lin et al.\ 2006; McCarthy et al.\ 
2007).  It is therefore interesting to see whether this 
experiment exposes any significant differences with respect 
to the results of our default merger simulation.

The dotted magenta curve in Figure~\ref{orbit_test} 
represents the final \flash to \gadget $K(<M_{\rm gas})$ 
ratio for the case where a dark matter only cluster merges 
with another cluster composed of both gas and dark matter.  
The results of this test are remarkably similar to that of 
our default merger case.  This indicates that the mechanism 
responsible for the difference in heating in the mesh and 
SPH simulations in the default merger simulation is also 
operating in this setup.  Although this does not pin down the 
difference between the mesh and SPH simulations, it does 
suggest that the difference has little to do with differences 
in the properties of the large hydrodynamic shock that 
occurs at core collision, as there is no corresponding large 
hydrodynamic shock in the case where one cluster is composed 
entirely of dark matter.  However, it is clear from 
Figure 3 that the difference between the default mesh and SPH 
simulations is established around the time of core 
collision, implying that some source of heating other than 
the large hydrodynamic shock is operating at this time (at 
least in the \flash simulation).  We return to this point in 
\S 4.

%-----------------------------------------------------------------------------

\section{What Causes The Difference?}

There are fundamental differences between Eulerian mesh-based 
and Lagrangian particle-based codes in terms of how they 
compute the hydrodynamic and gravitational forces.  Ideally, 
in the limit of sufficiently high resolution, the two 
techniques would yield identical results for a given initial 
setup.  Indeed, both techniques have been shown to match 
with high accuracy a variety of test problems with known 
analytic solutions.  However, as has been demonstrated above 
(and in other recent studies; e.g., Agertz et al.\ 2007; 
Trac et al.\ 2007; Wadsley et al.\ 2008) differences that 
do not appear to depend on resolution present themselves in 
certain complex, but astrophysically-relevant, circumstances.

In what follows, we explore several different possible 
causes for why the central heating that takes place in mesh 
simulations exceeds that in the SPH simulations.  The 
possible causes we explore include:

\begin{itemize}

\item{\S~\ref{gravitysolvers} A difference in gravity 
solvers - Most currently popular mesh codes (including 
\flash and \enzo) use a particle-mesh (PM) approach to 
calculate the gravitational force.  To accurately capture 
short range forces it is therefore necessary to have a 
finely-sampled mesh.  By contrast, particle-based codes 
(such as \gadget and \gasoline) often make use of tree 
algorithms or combined tree particle-mesh (TreePM) 
algorithms, where the tree is used to compute the short range 
forces and a mesh is used to compute long range forces.  
Since the gravitational potential can vary rapidly during 
major mergers and large quantities of mass can temporarily be 
compressed into small volumes, it is conceivable 
differences in the gravity solvers and/or the adopted 
gravitational force resolution could give rise to different 
amounts of entropy generation in the simulations.}

\item{\S~\ref{galileaninv} Galilean non-invariance of 
mesh codes - Given explicit dependencies in the Riemann 
solver's input states, all Eulerian mesh codes are 
inherently not Galilean invariant to some degree.  This can 
lead to spurious entropy generation in the cores of systems 
as they merely translate across the simulation volume (e.g., 
Tasker et al.\ 2008).}

\item{\S~\ref{viscosity} `Pre-shocking' in the SPH 
runs - 
Artificial viscosity is required in SPH codes to capture the 
effects of shock heating.  However, the artificial viscosity 
can in principle lead to entropy production in regions where 
no shocks should be present (e.g., Dolag et al.\ 2005).  If 
such pre-shocking is significant prior to core collision in 
our SPH simulations, it could result in a reduced efficiency 
of the primary shock.}

\item{\S~\ref{mixing} A difference in the amount of mixing in 
SPH and mesh codes - Mixing will be suppressed in 
standard SPH implementations where steep density gradients 
are present, since Rayleigh-Taylor and Kelvin-Helmholtz 
instabilities are artificially damped in such circumstances 
(e.g., Agertz et al.\ 2007).  In addition, the standard 
implementation of artificial viscosity will damp out even 
{\it subsonic} motions in SPH simulations, thereby inhibiting 
mixing (Dolag et al.\ 2005).  On the other hand, one expects 
there to be some degree of over-mixing in mesh codes, since 
fluids are implicitly assumed to be fully mixed on scales 
smaller than the minimum cell size.}
\\
\\
We now investigate each of these possible causes in turn.  We 
do not claim that these are the only possible causes for the 
differences we see in the simulations.  They do, however, 
represent the most commonly invoked possible solutions 
(along with hydrodynamic resolution, which we explored in \S 
3) to the entropy core discrepancy between SPH and mesh 
codes.

\end{itemize}

\subsection{Is it due to a difference in the gravity solvers?}
\label{gravitysolvers}

\begin{figure}
\centering
\includegraphics[width=8.4cm]{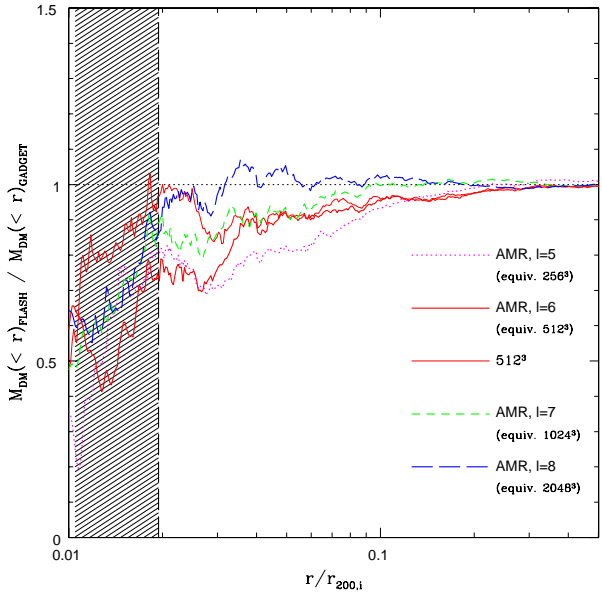}
\caption{A plot comparing the resulting dark matter mass
distributions for the default merger setup at 10 Gyr.  The dark matter 
mass profiles have been normalised to the final dark matter 
mass profile of the default resolution \gadget run.  The 
dotted magenta, solid red, short-dashed green, and 
long-dashed blue curves represent the \flash AMR runs 
with $l =$ 5, 6, 7, and 8, respectively, which
correspond to peak grid cell sizes of $\approx 78$, $39$,
$19.5$, and $9.8$ kpc (respectively).  The thin solid red 
curve represents the uniform $512^3$ \flash run.  The 
gravitational softening length adopted for the \gadget run is 
10 kpc.  For reference, $r_{200,i} \simeq 2062$ kpc.  The 
vertical dashed line indicates four softening lengths. The
\flash dark matter distribution converges to the \gadget 
result when the numerical resolutions become similar: the 
observed differences in gas entropy are not due to differences 
in the underlying dark matter dynamics.
}
\label{DMconverge}
\end{figure}

In the \flash simulations, gravity is computed using a 
standard particle-mesh approach.  With this approach,
the gravitational force will be computed accurately only 
on scales larger than the finest cell size.  
By contrast, the \gadget simulations make use of a combined 
TreePM approach, where the tree algorithm computes the short 
range gravitational forces and the particle-mesh algorithm is 
used only to compute long range forces.  To test whether or 
not differences in the gravity solvers (and/or gravitational 
force resolution) are important, we compare the final 
mass distributions of the dark matter in our simulations.  The 
distribution of the dark matter should be insensitive to the 
properties of the diffuse baryonic component, since its 
contribution to the overall mass budget is small by 
comparison to the dark matter.\footnote{We have explicitly 
verified this by running a merger between clusters with gas 
mass fractions that are a factor of 10 lower than assumed in 
our default run.}  Thus, the final distribution of the dark 
matter tells us primarily about the gravitational interaction alone 
between the two clusters.

Figure~\ref{DMconverge} shows the ratio of the final \flash 
dark matter mass profiles to the final 
\gadget dark matter mass profile. Recall that in all runs
the number of dark matter particles is the same. The differences
that are seen in this figure result from solving for
gravitational potential on a finer mesh. For the
lowest resolution \flash run, we see that the final 
dark matter mass profile deviates 
significantly from that of the default \gadget run for $r 
\la 0.04 r_{200,i}$.  However, this should not be surprising, 
as the minimum cell size in the default \flash run is $\sim 
0.02 r_{200,i}$.  By increasing the maximum refinement level, 
$l$, we see that the discrepancy between the 
final \flash and 
\gadget dark matter mass profiles is limited to smaller and 
smaller radii.  With $l = 8$, the minimum cell 
size is equivalent to the gravitational softening length 
adopted in the default \gadget run.  In this case, the 
final dark matter mass distribution agrees with that of the 
default \gadget run to within a few percent at all radii 
beyond a few softening lengths (or a few cell sizes), which is 
all that should be reasonably expected.  A comparison of the various 
\flash runs with one another (compare, e.g., the default \flash run 
with the $l=8$ run, for which there is a $\sim6$\% discrepancy out as far 
as $0.1 r_{200}$) may suggest a somewhat slower rate of convergence to the 
default \gadget result than one might naively have expected.  Given that 
we have tested the new FFT gravity solver against both the default
multigrid solver and a range of simple 
analytic problems and confirmed its accuracy to a much higher level than 
this, we speculate that the slow rate of convergence is due to the 
relatively small number of dark matter particles used in the mesh 
simulations.  In the future, it would be useful to vary the number of dark 
matter particles in the mesh simulations to verify this hypothesis.

In summary, we find that the resulting dark matter distributions 
agree very well in the \gadget and \flash simulations when the 
effective resolutions are comparable.  The intrinsic 
differences between the solvers therefore appear to be minor.  
More importantly for our purposes, even though the gravitational force 
resolution for the default \flash run is not as high as for 
the default \gadget run, this has no important consequences 
for the comparison of the final entropy distributions of the 
gas.  It is important to note that, even though 
the final mass distribution in the \flash simulations
shows small differences between $l = 6$ and 8, 
Figure~\ref{entropy_radius} shows that the entropy 
distribution is converged for $l \ge 6$
and is not at all affected by the improvement in the 
gravitational potential.

\subsection{Is it due to Galilean non-invariance of grid 
hydrodynamics?}
\label{galileaninv}

\begin{figure}
\centering
\includegraphics[width=8.4cm]{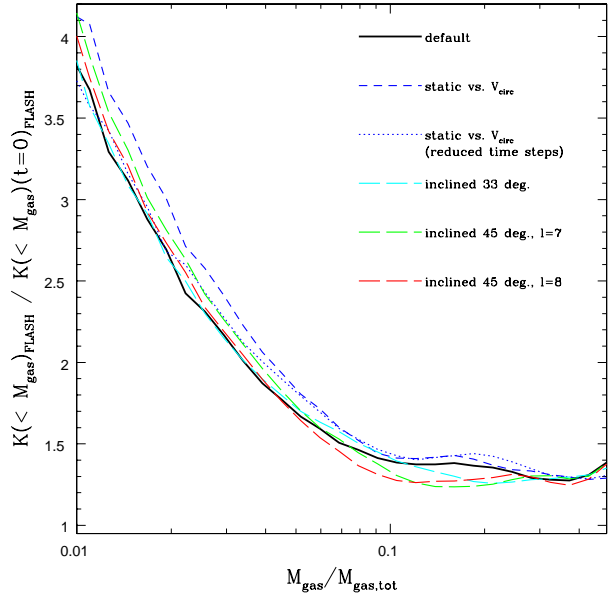}
\caption{Testing the effects of Galilean non-invariance on
the
\flash merger simulations.  Plotted is the final entropy
distribution, normalised to the initial one, for the default
\flash merger simulation and various different `takes' on
the default run.  The solid black curve represents the
default
run, the short-dashed blue curves represents a merger where
one cluster is held static and the other given a bulk
velocity
twice that in the default run (i.e., the relative velocity
is
unchanged from the default run), and the dotted blue curve
represents this same merger but with the size of the time
steps reduced by an order of magnitude.  The dashed cyan,
green, and red lines represent mergers that take place on an
oblique angle to mesh at 33 degrees, 45 degrees with $l=7$ 
and 45 degrees with $l=8$,
respectively.  This comparison illustrates that the
effects of Galilean non-invariance on the resulting entropy
distribution are minor and do not account for the difference
in the entropy core amplitudes of the mesh and SPH
simulations.
}
\label{flashgalilean}
\end{figure}

Due to the nature of Riemann solvers (which are a fundamental 
feature of AMR codes), it is possible for the evolution of a 
system to be Galilean non-invariant.  This arises from the 
fixed position of the grid relative to the fluid. The Riemann 
shock tube initial conditions are constructed by determining 
the amount of material that can influence the cell boundary 
from either side, within a given time step based on the sound 
speed. The Riemann problem is then solved at 
the boundary based on the fluid properties either side of the boundary. 
By applying a bulk velocity to the medium in a given direction, the 
nature of the solution changes. Although an ideal solver
would be able to decouple the bulk velocity from the velocity 
discontinuity at the shock, the discrete nature of the problem 
means that the code may not be Galilean invariant. Since one expects 
large bulk motions to be relevant for cosmological structure 
formation, and clearly is quite relevant for our merger 
simulations, it is important to quantify what effects (if any) 
Galilean non-invariance has on our AMR simulations.

We have tested the Galilean non-invariance of our \flash 
simulations in two ways.  In the first test, we simulate an 
isolated cluster moving across the mesh with an (initial) bulk 
velocity of $V_{\rm circ}(r_{200})$ ($\simeq 1444$ km/s) and 
compare it to an isolated cluster with zero bulk velocity.  
This is similar to the test carried out recently by Tasker 
et al.\ (2008).  In agreement with Tasker et al.\ (2008), we 
find that there is some spurious generation of entropy in 
the very central regions ($M_{\rm gas}/M_{\rm gas,tot} \la 
0.03$) of the isolated cluster that was given an initial 
bulk motion.  However, after $\approx 2$ Gyr of evolution 
(i.e., the time when the clusters collide in our default 
merger simulation), the increase in the central entropy is 
only $\sim 10$\%.  This is small in comparison to the 
$\sim300$\% jump that takes place at core collision in 
our merger simulations.  This suggests that spurious 
entropy generation prior to the merger is minimal and 
does not account for the difference we see between the SPH and 
AMR simulations.

In the second test, we consider different implementations of 
the default merger simulation.  In one case, instead of giving
both systems equal but opposite bulk velocities (each with
magnitude $0.5 V_{\rm circ}$), we fix one and give the other
an initial velocity that is twice the default value, so that
the relative velocity is unchanged.  (We also tried reducing
the size of the time steps for this simulation by an order
magnitude.)  In addition, we have tried mergers that take
place at oblique angles relative to the grid.  If the merger 
is well-resolved and the dynamics are Galilean invariant, 
all these simulations should yield the same result.

Figure~\ref{flashgalilean} shows the resulting entropy 
distributions for these different runs.  The results of this 
test confirm what was found above; i.e., that there is some 
dependence on the reference frame adopted, but that this 
effect is minor in general (the central entropy is modified 
by $\la 10$\%) and does not account for the discrepancy we 
see between entropy core amplitudes in the default \gadget 
and \flash simulations.

\begin{figure}
\centering
\includegraphics[width=8.4cm]{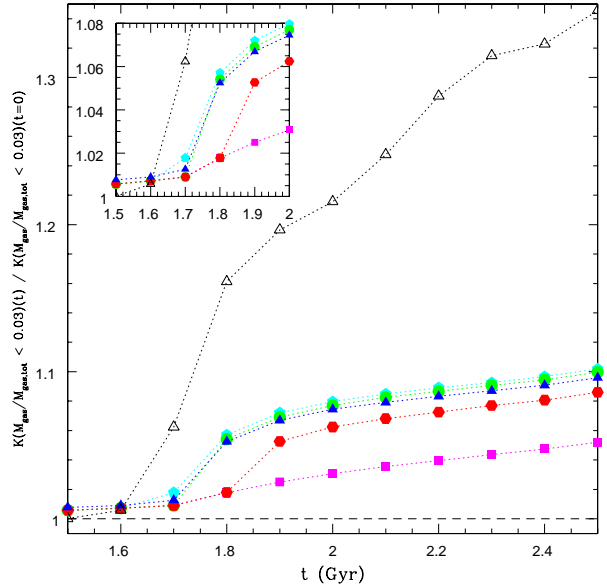}
\caption{Testing the effects of pre-shocking due to
artificial
viscosity in the default \gadget merger simulation.  This
plot shows the evolution of the central entropy (enclosing 3\% of
the gas mass) around the time of first core collision.  The
solid blue triangles represent the default \gadget simulation.  The
solid cyan points, solid green points, and solid red points represent
runs where the artificial viscosity is kept at a very low level
($\alpha_{\rm visc} = 0.05$) until $t \approx$ 1.6, 1.7,
1.8 Gyr, respectively, at which point the artificial viscosity
is set back to its default value.  The solid magenta squares
represent a run with low artificial viscosity throughout,
and the open triangles represent the default \flash
simulation. Reducing the value of the artificial viscosity parameter
before the cores collide delays the increase in entropy (cyan, green
and solid red), however as soon as the original value is restored, the
entropy $K$ increases to a level nearly independent of when $\alpha$
was restored. Therefore pre-shocking has little effect on the
post-shock value of $K$.
}
\label{viscosityswitch}
\end{figure}

\subsection{Is it due to `pre-shocking' in SPH?}
\label{viscosity}

Artificial viscosity is required in SPH codes in order to 
handle hydrodynamic shocks.  The artificial viscosity acts 
as an excess pressure in the equation of motion, converting 
gas kinetic energy into internal energy, and therefore raising 
the entropy of the gas.  In standard SPH implementations, the 
magnitude of the artificial viscosity is fixed in both space 
and time for particles that are approaching one another (it 
being set to zero otherwise).  This implies that even in cases 
where the Mach number is less than unity, i.e., where formally 
a shock should not exist, (spurious) entropy generation can 
occur.  This raises the possibility that significant 
`pre-shocking' could occur in our SPH merger simulations.  
This may have the effect of reducing the efficiency of the 
large shock that occurs at core collision and could therefore 
potentially explain the discrepancy between the mesh and SPH 
simulations.

\begin{figure*}
\centering
\leavevmode
\epsfysize=8.4cm \epsfbox{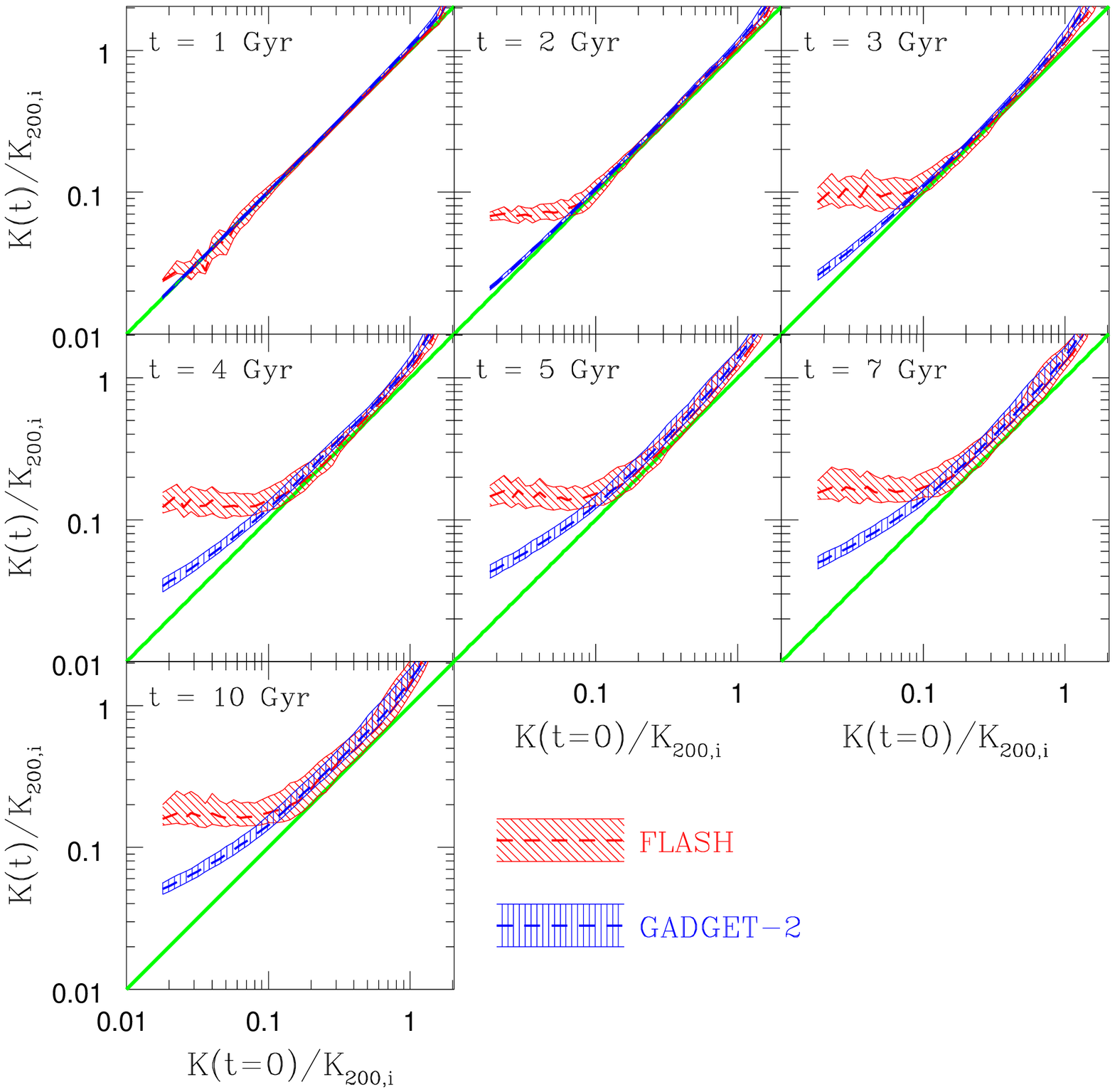}
\epsfysize=8.4cm \epsfbox{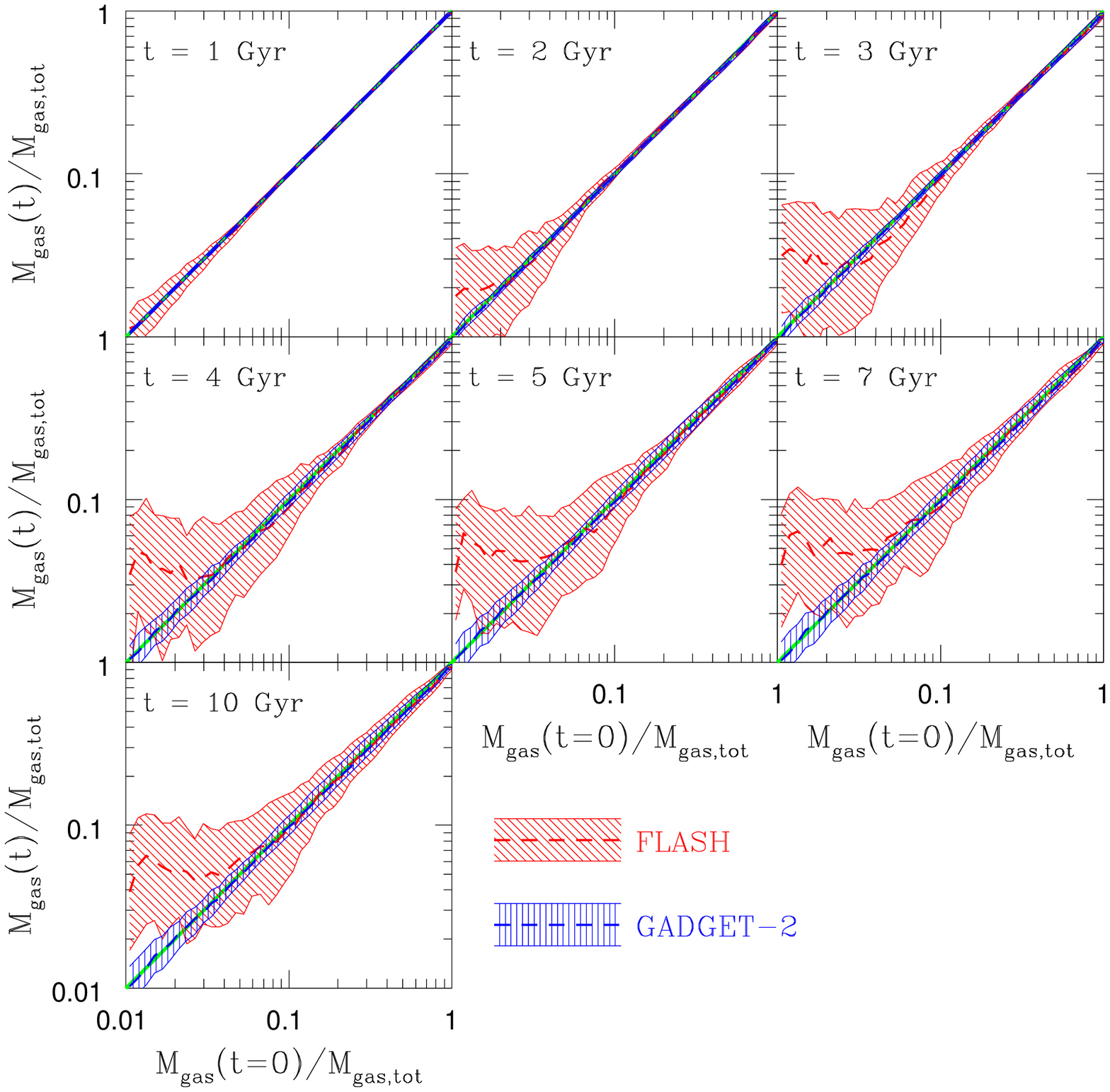}
\caption{Quantifying the amount of heating and mixing in the
default \gadget and \flash merger simulations.  {\it Left:}
The entropy of particles (tracer particles in the case of
\flash) at time $t$ vs. the initial entropy of those
particles.  The solid green line is the line of equality
[$K(t) = K(t=0)$; i.e., no heating].  The
shaded blue and red regions represent the distributions
from the \gadget and \flash simulations, respectively.  They
enclose 50\% of the particles; i.e., the lower/upper bounds
represent the 25th/75th percentiles for $K(t)$ at
fixed $K(t=0)$.  The dashed blue and red lines
represent the median $K(t)$ at fixed $K(t=0)$. The central
entropy in the \flash runs increases significantly more than 
in the \gadget run when the cores collide, at $t\sim 2$~Gyr, 
the increase in entropy later is similar between the two 
codes. The scatter in \flash entropy is also much larger than 
in \gadget. {\it Right:} The enclosed gas mass of particles 
(tracers particles in the case of \flash) at time $t$ vs. the 
initial enclosed mass of those particles.  The enclosed gas 
mass of each particle is calculated by summing the masses of
all other particles (or cells) with entropies lower than the
particle under consideration.  The solid green line is the 
line of equality [$M_{\rm gas}(t) = M_{\rm gas}(t=0)$; i.e., 
no mass
mixing].  The shaded blue and red regions represent the
distributions from the \gadget and \flash simulations,
respectively.  They enclose 50\% of the particles; i.e., the
lower/upper bounds represent the 25th/75th percentiles for
$M_{\rm gas}(t)$ at fixed $M_{\rm gas}(t=0)$.  The dashed
blue and red lines represent the median $M_{\rm gas}(t)$ at
fixed $M_{\rm gas}(t=0)$.  Particles in \flash mix much more
than in \gadget.}
\label{tracers}
\end{figure*}

Dolag et al.\ (2005) raised this possibility and tested it in 
SPH cosmological simulations of massive galaxy clusters.  
They implemented a new variable artificial viscosity scheme 
by embedding an on-the-fly shock detection algorithm in 
\gadget that indicates if particles are in a supersonic flow
or not. If so, the artificial viscosity is set to a typical 
value, if not the artificial viscosity is greatly reduced.  
This new implementation should significantly reduce the 
amount of pre-shocking that takes place during formation of 
the clusters.  The resulting clusters indeed had somewhat 
higher central entropies relative to clusters 
simulated with the standard artificial viscosity 
implementation (although the new scheme does not appear to 
fully alleviate the discrepancy between mesh and SPH 
codes).  However, whether the central entropy was raised 
because of the reduction in pre-shocking or if it was due 
to an increase in the amount of mixing is unclear.

Our idealised mergers offer an interesting opportunity to 
re-examine this test.  In particular, because of the 
symmetrical geometry of the merger, little or no mixing is 
expected until the cores collide, as prior to this time there 
is no interpenetration of the gas particles belonging to the 
two clusters (we have verified this).  This means that we are 
in a position to isolate the effects of pre-shocking 
from mixing early on in the simulations.  To do so, we have 
devised a crude method meant to mimic the variable 
artificial viscosity scheme of Dolag et al.\ (2005).  In 
particular, we run the default merger with a low artificial 
viscosity (with $\alpha_{\rm visc} = 0.05$, i.e., 
approximately the minimum value adopted by Dolag et al.\ ) 
until the cores collide, at which point we switch the 
viscosity back to its default value.  We then examine the 
amount of entropy generated in the large shock.

Figure~\ref{viscosityswitch} shows the evolution of the 
central entropy around the time of core collision.  Shown are 
a few different runs where we switch the artificial 
viscosity back to its default value at different times (since 
the exact time of `core collision' is somewhat ill-defined).  
Here we see that prior to the large shock very little entropy 
has been generated, which is expected given the low 
artificial viscosity adopted up to this point.  A comparison 
of these runs to the default \gadget simulation (see inset in
Figure~\ref{viscosityswitch}) shows that there is evidence for  a small 
amount of pre-shocking in the default run.
However, we find that for the cases where the artificial viscosity is
set to a low value, the resulting entropy jump (after the viscosity is
switched back to the default value) is nearly the same as in the
default merger simulation. In other words, pre-shocking appears to 
have had a minimal effect on the strength of the heating that 
occurs at core collision in the default SPH simulation. This argues 
against pre-shocking as the cause of the difference we see between 
the mesh and SPH codes.

Lastly, we have also tried varying $\alpha_{\rm visc}$ 
over the range $0.5$ and $1.0$ (i.e., 
values typically adopted in SPH studies; Springel 2005b) for the 
default \gadget run.  We find that the SPH results are robust to 
variations in $\alpha_{\rm visc}$ and cannot reconcile the 
differences between SPH and AMR results.

\begin{figure*}
\centering
\leavevmode
\epsfysize=8.4cm \epsfbox{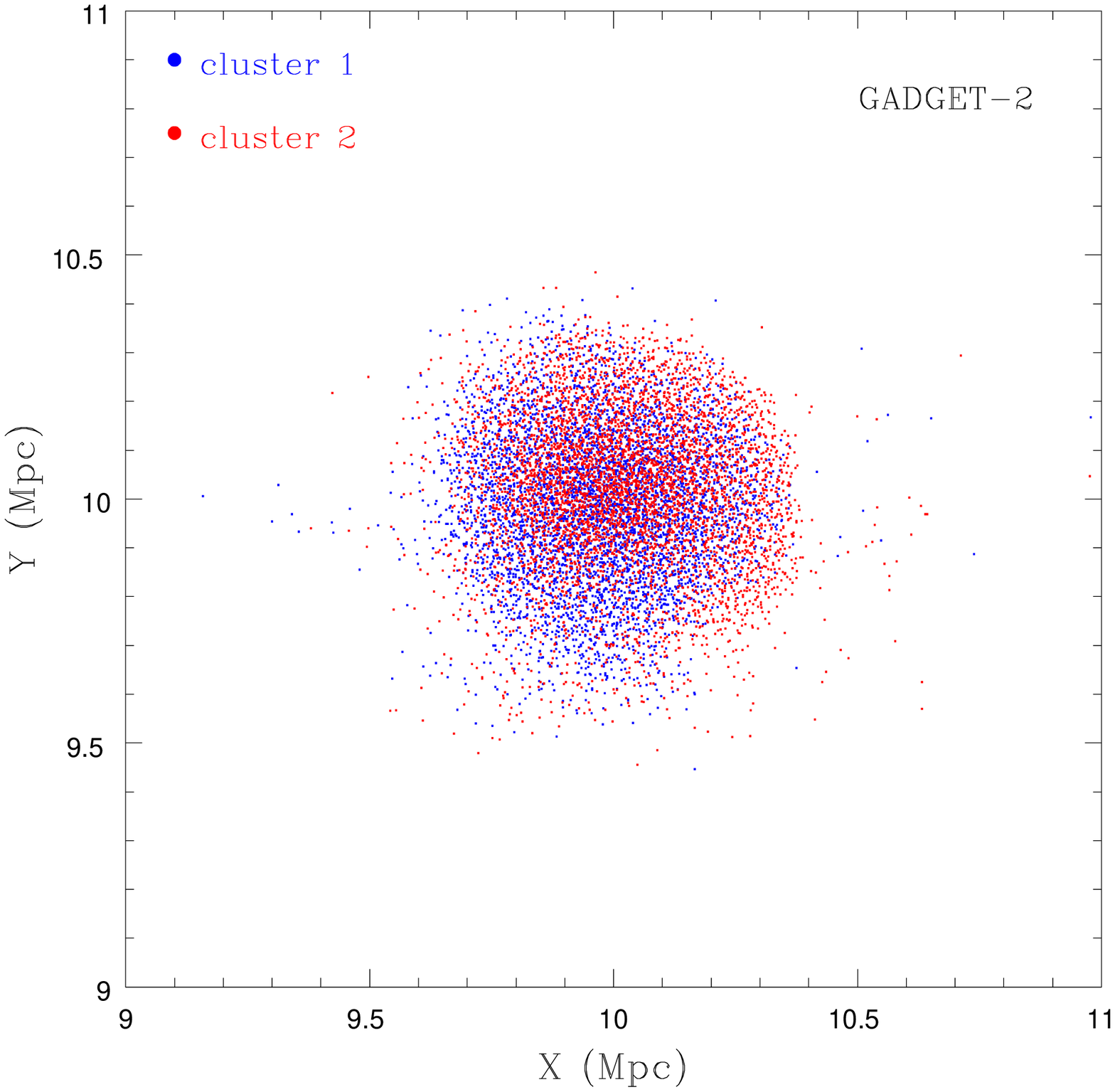}
\epsfysize=8.4cm \epsfbox{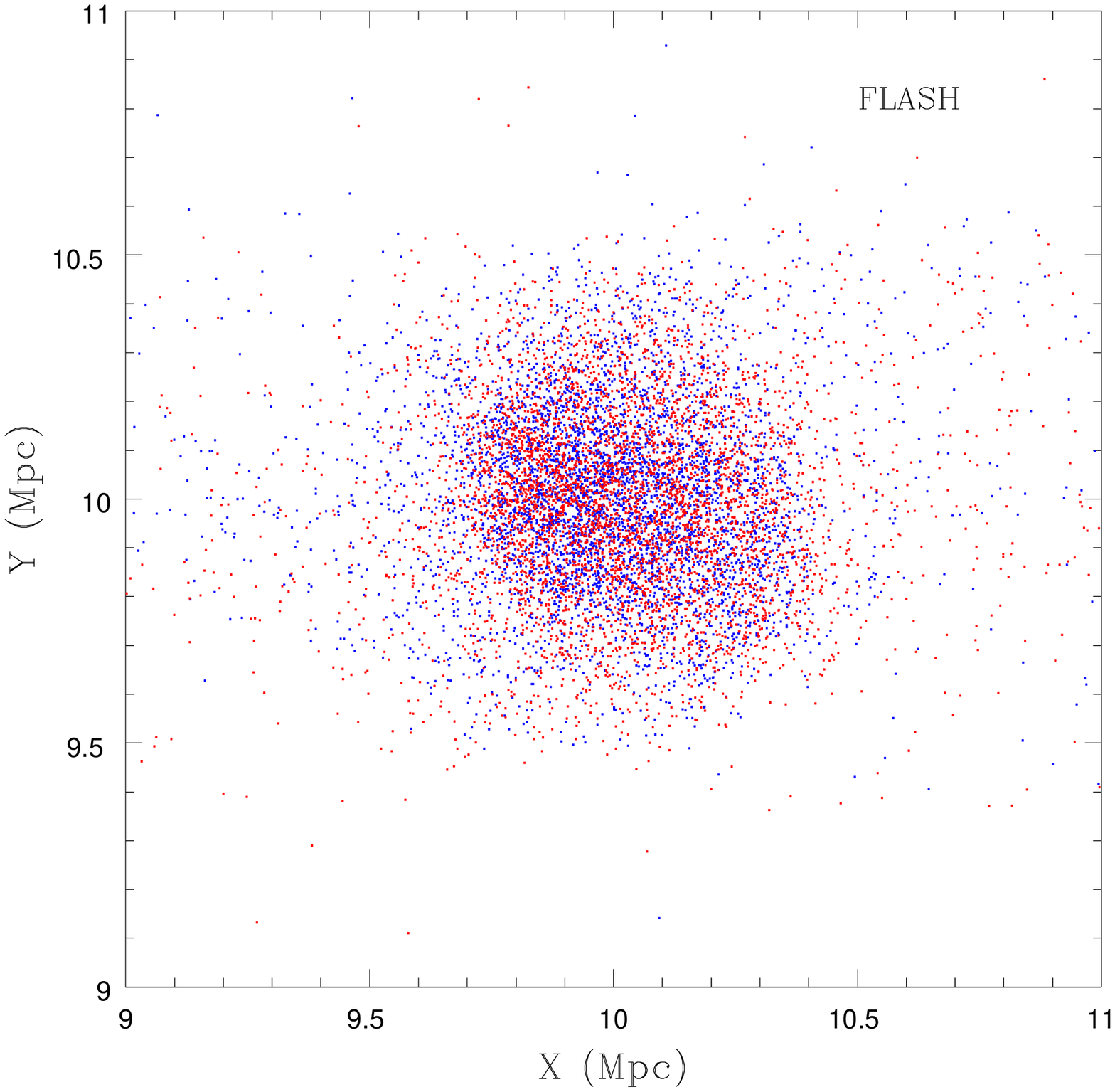}
\caption{The final spatial distribution of particles
(tracer particles in the case of \flash) with the lowest
initial entropies (we select the central 5\% of
particles/tracer particles in both clusters).  The blue
points represent particles belonging to one of the clusters and the
red points represent particles belonging to the other.  {\it
Left:} The low resolution \gadget simulation.  {\it Right:} The
default \flash simulation.  There is a high degree of
mixing in the mesh simulation, whereas there remain two 
distinct blobs corresponding to the infallen clusters in the SPH
simulation. The difference arises immediately following core collision
through the turbulent mixing that it drives.
}
\label{particles}
\end{figure*}

\subsection{Is it due to a difference in the amount of 
mixing in SPH and mesh codes?}
\label{mixing}

Our experiments with off-axis collisions and collisions with
a cluster containing only dark matter suggest that mixing
plays an important role in generating the differences between
the codes. 
Several recent studies (e.g., Dolag et al.\ 2005; Wadsley et 
al.\ 2008) have argued that mixing is handled poorly in 
standard implementations of SPH, both because (standard) 
artificial viscosity acts to damp turbulent motions and 
because the growth of KH and RT instabilities is inhibited in 
regions where steep density gradients are present (Agertz et 
al.\ 2007).  Using cosmological SPH simulations that have been 
modified in order to enhance mixing\footnote{Note the 
that modifications implemented by Dolag et al.\ (2005) and 
Wadsley et al.\ (2008) differ.  As described in \S 4.3, 
Dolag et al.\ (2005) implemented a variable artificial 
viscosity, whereas Wadsley et al.\ (2008) introduced a 
turbulent heat flux term to the Lagrangian energy 
equation in an attempt to explicitly model turbulent 
dissipation.}, Dolag et al.\ (2005) and Wadsley et al. (2008)
have shown that it is possible to generate higher central 
entropies in their galaxy clusters (relative to clusters 
simulated using standard implementations of SPH), yielding 
closer agreement with the results of cosmological mesh 
simulations.  This is certainly suggestive that mixing may be 
the primary cause of the discrepancy between mesh and SPH 
codes.  However, these authors did not run mesh 
simulations of galaxy clusters and therefore did not perform 
a direct comparison of the amount of mixing in SPH vs. mesh 
simulations of clusters.  Even if one were to directly 
compare cosmological SPH and mesh cluster simulations, the 
complexity of the cosmological environment and the 
hierarchical growth of clusters would make it difficult to 
clearly demonstrate that mixing is indeed the difference.

Our idealised mergers offer a potentially much cleaner way to 
test the mixing hypothesis.  To do so, we re-run the default 
\flash merger simulation but this time we include a large 
number of `tracer particles', which are massless and follow 
the hydrodynamic flow of the gas during the simulation.  
The tracer particles are advanced using a second order 
accurate predictor-corrector time advancement scheme with the 
particle velocities being interpolated from the grid (further
details are given in the flash
manual (version 2.5) at: \\ http://flash.uchicago.edu/).
Each tracer particle has a unique ID that is preserved 
throughout the simulation, allowing us to track 
the gas in a Lagrangian fashion, precisely as is done in 
Lagrangian SPH simulations.  To simplify the comparison 
further, we initially distribute the tracer particles within the 
two clusters in our \flash simulation in exactly the same way 
as the particles in our initial \gadget setup.

In the left hand panel of Figure~\ref{tracers}, we plot the 
final vs. the initial entropy of particles in the default \gadget and 
\flash merger simulations.  This plot clearly 
demonstrates that the lowest-entropy gas is preferentially 
heated in both simulations, however the degree of heating of 
that gas in the mesh simulation is much higher than in the SPH 
simulation.  Consistent with our analysis in \S 3, we find 
that the bulk of this difference is established around the 
time of core collision.  It is also interesting that the 
scatter in the final entropy (for a given initial entropy) 
is much larger in the mesh simulation.  The larger scatter 
implies that convective mixing is more prevalent in 
the mesh simulation. At or immediately following core 
collision ($t \approx 2-3$ Gyr), there is an indication 
that, typically, gas initially at the very centre of the two 
clusters (which initially had the lowest entropy) has been 
heated more strongly than gas further out [compare, e.g., 
the median $K(t=5 {\rm Gyr})$ at $K(t=0)/K_{200,i} \approx 
0.02$ to the median at $K(t=0)/K_{200,i} \approx 0.08$].  
Such an entropy inversion does not occur in the SPH 
simulations and likely signals that the extra mixing in the 
mesh simulation has boosted entropy production. 

\begin{figure*}
\centering
\leavevmode
\epsfysize=8.4cm \epsfbox{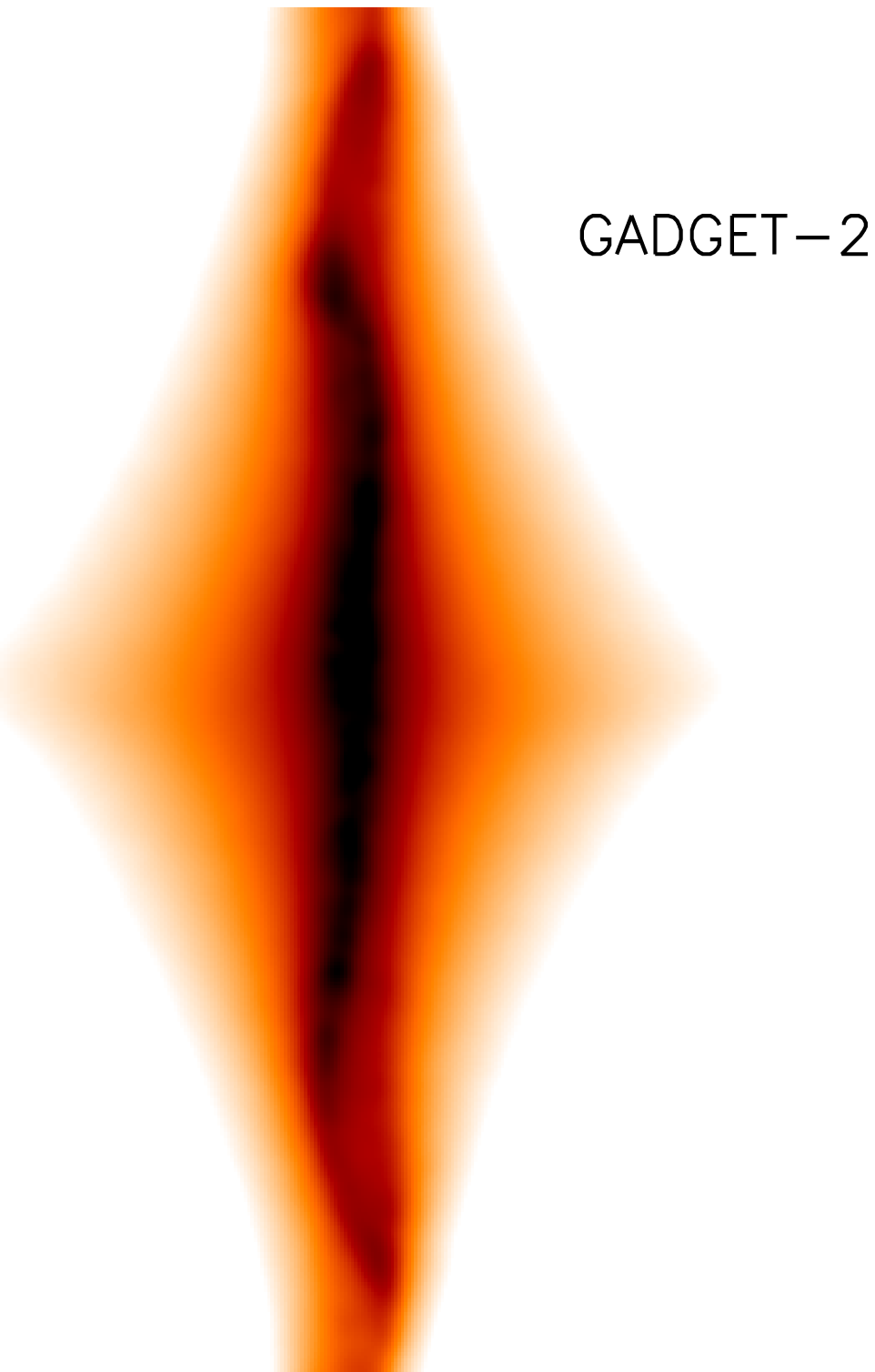}
\epsfysize=8.4cm \epsfbox{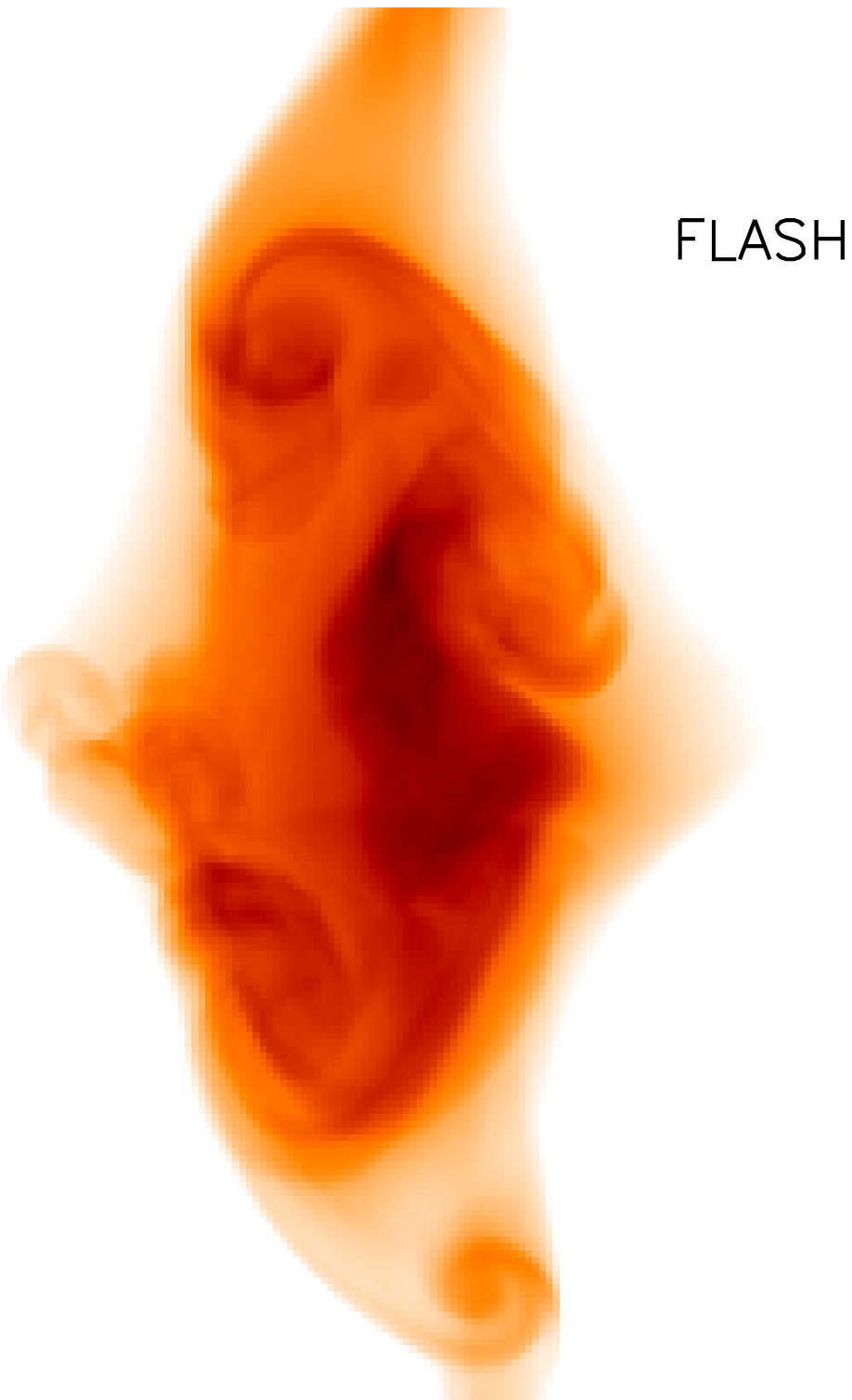}
\caption{Logarithmic projected entropy maps of the default
\gadget simulation and \flash simulation with $l=8$ at $t = 
2.3$ Gyr, just after the collision of the 
cores.  Note that the peak spatial resolutions of the two simulations are
similar (approx.\ 10 kpc; we also note that the {\it median} SPH 
smoothing length for the default \gadget run is $\approx 20$ kpc within 
the mixing region, $r \la 200$ kpc.)  To highlight the central regions, 
we have reset the 
value of any pixel with projected entropy greater than $0.5 
K_{200,i}$ to $0.5 K_{200,i}$.  In
these maps, the minimum entropy (black) is $\approx 0.07
K_{200,i}$ and the maximum entropy (white) is $0.5
K_{200,i}$.  The maps are 2~Mpc on a side, and project over
a depth of 2~Mpc. Note that the images are not directly comparable
with figure 2, where a slice is shown. {\it Left:}
the \gadget simulation.  {\it Right:} the \flash
simulation.  The \flash entropy distribution is characterised 
by vortices on a range of scales, which mix gas with 
different entropies. These vortices are mostly absent in the 
\gadget simulation.}
\label{instabilities}
\end{figure*}

In the right hand panel of Figure~\ref{tracers} we plot the
final vs. the initial enclosed gas mass of particles in the 
default \gadget and \flash merger simulations.  The enclosed 
gas mass of each particle (or tracer particle) is calculated 
by summing the masses of all other particles (or cells) with 
entropies lower than the particle under 
consideration\footnote{In convective equilibrium, the 
enclosed gas mass calculated in this way also corresponds 
to the total mass of gas of all other particles (or cells) 
within the cluster-centric radius (or at lower, more negative 
gravitational potential energies) of the particle under 
consideration.  We have verified this for the final output 
when the merged system has relaxed by, instead of 
summing the masses of all particles with entropy lower than $K_i$, 
by summing the masses of all particles with potentials lower 
than $\Phi_i$ for the $i$th particle.}.  This plot confirms 
our mixing expectations based on the entropy plot in the 
left hand panel.  In particular, only a small amount of mass 
mixing is seen in the SPH simulation, whereas in the mesh 
simulation the central $\sim 5\%$ of the gas mass has been 
fully mixed.

The higher degree of mixing in the \flash simulation is shown 
pictorially in Figure~\ref{particles}.  The left panel shows 
the final spatial distribution of the initially 
lowest-entropy particles in \gadget simulation, while the 
right panel is the analogous plot for tracer particles in 
the \flash simulation (see figure caption).  The larger 
degree of mixing in the mesh simulation relative to the SPH 
simulation is clearly evident. In the \flash simulation, particles
from the two clusters are intermingled in the final state,
while distinct red and blue regions are readily apparent in
the SPH calculation, a difference which arises immediately
following core collision.

The increased mixing boosts entropy production in the \flash 
simulations, but what is the origin of the increased 
mixing?  We now return to the point raised in \S 3, that the 
bulk of the difference between mesh and SPH simulation is 
established around the time of core collision.  This is in 
spite of the fact that significant entropy generation 
proceeds in both simulations until $t \sim 6$ Gyr. Evidently, 
both codes treat the entropy generation in the re-accretion 
phase in a very similar manner.  What is different about the 
initial core collision phase?  As pointed out recently by 
Agertz et al.\ (2007), SPH suppresses the growth of 
instabilities in regions where steep density gradients 
are present due to spurious pressure forces acting on the 
particles.  Could this effect be responsible for the 
difference we see?  To test this idea, we generate 2D 
projected entropy maps of the SPH and mesh simulations to 
search for signs of clear instability development.  In the 
case of the SPH simulation, we first 
smooth the particle entropies (and densities) onto a 3D grid 
using the SPH smoothing kernel and the smoothing lengths of 
each particle computed by \gadget.  We then compute a 
gas mass-weighted projected entropy by projecting along the 
$z$-axis.  In the case 
of the \flash simulation, the cell entropies and densities 
are interpolated onto a 3D grid and projected in the same 
manner as for the \gadget simulation.  

Figure~\ref{instabilities} shows a snap shot of the two simulations
at $t = 2.3$ Gyr, just after core collision. Large 
vortices and eddies are easily visible in the projected entropy 
map of the \flash simulation but none are evident in the 
\gadget simulation. In order to study the duration of
these eddies, we have generated 100 such snap shots 
for each simulation, separated by fixed 0.1 Gyr intervals. Analysing the 
projected entropy maps as a movie\footnote{For movies see {\em 
``Research: Cores in Simulated Clusters'' } at http://www.icc.dur.ac.uk/}, we
find that these large vortices and eddies persist in the \flash 
simulation 
from $t \approx 1.8 - 3.2$ Gyr.  This corresponds very well with 
the timescale over which the difference between the SPH and mesh 
codes is established (see, e.g., the dashed black curve in the top 
right panel of Figure~\ref{entropygastime}). 

We therefore conclude that extra mixing in the mesh 
simulations, brought on by the growth of instabilities around 
the time of core collision, is largely responsible for the 
difference in the final entropy core amplitudes between the 
mesh and SPH simulations. Physically, one expects the development 
of such instabilities, since the KH timescale, $\tau_{\rm KH}$, is 
relatively short at around the time of core collision.
We therefore conclude that there is a degree of under-mixing in 
the SPH simulations\footnote{But we note that very 
high resolution 2D `blob' simulations carried out by Springel 
(2005b) do clearly show evidence for vortices.  It is presently 
unclear if these are a consequence of the very different 
physical setup explored in that study (note that the gas density 
gradients are much smaller than in the present study) or the 
extremely high resolution used in their 2D simulations, and 
whether or not these 
vortices lead to enhanced mixing and entropy production.} 
Whether or not the \flash simulations 
yield the correct result, however, is harder to ascertain.  
As fluids are {\it forced} to numerically mix on scales 
smaller than the minimum cell size, it is possible that there 
is a non-negligible degree of over-mixing in the mesh 
simulations.  Our resolution tests (see \S 3) show evidence 
for the default mesh simulation being converged, but it may 
be that the resolution needs to be increased by much larger 
factors than we have tried (or are presently accessible with 
current hardware and software) in order to see a difference.

\section{Summary and Discussion}

In this paper, we set out to investigate the origin of the 
discrepancy in the entropy structure of clusters formed in 
Eulerian mesh-based simulations compared to those formed in 
Lagrangian SPH simulations.  While SPH simulations form 
clusters with almost powerlaw entropy distributions down to 
small radii, Eulerian simulations form much larger cores with 
the entropy distribution being truncated at significantly 
higher values. Previously it has been suspected that this 
discrepancy arose from the limited resolution of the mesh 
based methods, making it impossible for such codes to 
accurately trace the formation of dense gas structures at
high redshift. 

By running simulations of the merging of idealised clusters, 
we have shown that this is not the origin of the discrepancy. 
We used the \gadget code (Springel 2005b) to compute the SPH 
solution and the \flash code (Fryxell et al.\ 2000) to 
compute the Eulerian mesh solution. In these idealised 
simulations, the initial gas density structure is resolved 
from the onset of the simulations, yet the final 
entropy distributions are significantly different. The 
magnitude of the difference generated in idealised mergers 
is comparable to that seen in the final clusters formed in 
full cosmological simulations.  A resolution study shows 
that the discrepancy in the idealised simulations cannot be 
attributed to a difference in the effective resolutions of 
the simulations.  Thus, the origin of the discrepancy must 
lie in the code's different treatments of gravity and/or 
hydrodynamics.

We considered various causes in some detail. We found that 
the difference was {\em not} due to:
\begin{itemize}
\item{The use of different gravity solvers. The two codes 
differ in that \gadget uses a TreePM method to determine 
forces, while \flash uses the PM method alone. The different 
force resolutions of the codes could plausibly lead to 
differences in the energy transfer between gas and dark 
matter. Yet we find that the dark matter distributions 
produced by the two codes are almost identical when the mesh 
code is run at comparable resolution to the SPH code.}
\item{Galilean non-invariance of mesh codes. We investigate 
whether the results are changed if we change the rest-frame 
defined by the hydrodynamic mesh. Although we find that an 
artificial core can be generated in this way in the mesh 
code, its size is much smaller than the core formed once the 
clusters collide, and is not enough to explain the difference 
between \flash and \gadget. We show that most of 
this entropy difference is generated in the space of $\sim 1$ 
Gyr when the cluster cores first collide.}
\item{Pre-shocking in SPH. We consider the possibility that 
the artificial viscosity of the SPH method might generate 
entropy in the flow prior to the core collision, thus 
reducing the efficiency with which entropy is
generated later. By greatly reducing the artificial 
viscosity ahead of the core collision, we show that this 
effect is negligible.}
\end{itemize}

Having shown that none of these numerical issues can explain 
the difference of the final entropy distributions, we 
investigated the role of fluid mixing in the two codes.  
Several recent studies (e.g., Dolag et al.\ 2005; Wadsley et 
al.\ 2008) have argued that if one increases the amount 
of mixing in SPH simulations the result is larger cluster 
entropy cores that resemble the AMR results.  While this is 
certainly suggestive, it does not clearly demonstrate that it 
is the enhanced mixing in mesh simulations that is indeed the 
main driver of the difference (a larger entropy core in the
mesh simulations need not necessarily have been established 
by mixing).  By injecting tracer particles into our \flash 
simulations, we have been able to make an explicit comparison 
of the amount of mixing in the SPH and mesh simulations of 
clusters.  We find very substantial differences. In the SPH 
computation, there is a very close relation between the 
initial entropy of a particle and its final entropy. In
contrast, tracer particles in the \flash simulation only 
show a close connection for high initial entropies. The 
lowest $\sim5$\% of gas (by initial entropy) is completely 
mixed in the \flash simulation. We conclude that mixing and
fluid instabilities are the cause of the discrepancy between 
the simulation methods.

The origin of this mixing is closely connected to the 
suppression of turbulence in SPH codes compared to the 
Eulerian methods. This can easily be seen by comparing the 
flow structure when the clusters collide: while the \flash 
image is dominated by large scale eddies, these are absent
from the SPH realisation (see Figure~\ref{instabilities}). It 
is now established that SPH codes tend to suppress the 
growth of Kelvin-Helmholtz instabilities in shear flows, and 
this seems to be the origin of the differences in our 
simulation results (e.g., Agertz et  al.\ 2007).  These 
structures result in entropy generation through mixing, an 
irreversible process whose role is underestimated by the SPH 
method.  Of course, it is not clear that the turbulent 
structures are correctly captured in the mesh simulations 
(Iapichino \& Niemeyer 2008; Wadsley et al.\ 2008). The mesh 
forces fluids to be mixed on the scale of individual cells. 
In nature, this is achieved through turbulent cascades that 
mix material on progressively smaller and smaller scales: the 
mesh code may well overestimate the speed and effectiveness 
of this process.  Ultimately, deep X-ray observations may be 
able to tell us whether the mixing that occurs in the 
mesh simulations is too efficient.  An attempt at 
studying large-scale turbulence in clusters was made recently 
by Schuecker et al.\ (2004).  Their analysis of {\it 
XMM-Newton} observations of the Coma cluster indicated the 
presence of a scale-invariant pressure fluctuation spectrum 
on scales of 40-90 kpc and found that it could be 
well described by a projected Kolmogorov/Oboukhov-type 
turbulence spectrum.  If the observed pressure fluctuations 
are indeed driven by scale-invariant turbulence, this would 
suggest that current mesh simulations have the resolution 
required to accurately treat the turbulent mixing process.
Alternatively, several authors have suggested that ICM may be highly 
viscous (eg., Fabian et al.\ 2003) with the result that fluid 
instabilities will be strongly suppressed by 
physical processes. This might favour the use of SPH methods which
include a physical viscosity (Sijacki \& Springel 2006).

It is a significant advance that we now understand the 
origin of this long standing discrepancy.  Our work also has 
several important implications.  Firstly, as outlined in \S 1, 
there has been much discussion in the recent literature on 
the competition between heating and cooling in galaxy 
groups and clusters.  The current consensus is that heating 
from AGN is approximately sufficient to offset cooling losses 
in observed cool core clusters (e.g., McNamara \& Nulsen 
2007).  However, observed present-day AGN power output seems 
energetically incapable of explaining the large number of 
systems that do not possess cool cores\footnote{Recent 
estimates suggest that $\sim50$\% of all massive X-ray 
clusters in flux-limited samples do not 
have cool cores (e.g., Chen et al.\ 2007).  Since at fixed 
mass cool core clusters tend to be more luminous than 
non-cool core clusters, the fraction of non-cool core 
cluster in flux-limited samples may actually be an 
underestimate of the true fraction.} (McCarthy et al.\ 2008).  
Recent high resolution X-ray observations demonstrate that 
these systems have higher central entropies than typical cool 
core clusters (e.g., Dunn \& Fabian 2008).  One way of 
getting around the energetics issue is to invoke an early 
episode of preheating (e.g., Kaiser 1991; Evrard \& Henry 1991).  
Energetically, it is more efficient to raise the entropy of 
the (proto-)ICM prior to it having fallen into the cluster 
potential well, as its density would have been much lower 
than it is today (McCarthy et al.\ 2008).  Preheating remains 
an attractive explanation for these systems.  

However, as we have seen from our idealised merger 
simulations, the 
amount of central entropy generated in our mesh simulations 
is significant and is even comparable to the levels observed 
in the central regions of non-cool core clusters.  It is 
therefore tempting to invoke mergers and the mixing they 
induce as an explanation for these systems.  However, before 
a definitive statement to this effect can be made, much 
larger regions of parameter space should be explored.  
In particular, a much larger range of impact parameters and 
mass ratios is required, in addition to switching on the 
effects of radiative cooling (which we have neglected in the 
present study).  This would be the mesh code analog of the 
SPH study carried out by Poole et al. (2006; see 
also Poole et al.\ 2008).  We leave this for future work. 
Alternatively, large cosmological mesh simulations, which 
self-consistently track the hierarchical growth of clusters, 
would be useful for testing the merger hypothesis.  Indeed, 
Burns et al.\ (2008) have recently carried out a large mesh 
cosmological simulation (with the \enzo code) and argue that 
mergers at high redshift play an important role in the 
establishment of present-day entropy cores.  However, these 
results appear to be at odds with the cosmological mesh 
simulations (run with the \art code) of Nagai et al.\ (2007) 
(see also Kravtsov et al.\ 2005).  These authors find that 
most of their clusters have large cooling flows at the 
present-day, similar to what is seen in some SPH cosmological 
simulations (e.g., Kay et al.\ 2004; Borgani et al.\ 2006).
On the other hand, the SPH simulations of Keres et al.\ (2008) 
appear to yield clusters with large entropy cores.  This may be ascribed to 
the lack of effective feedback in their simulations, as radiative cooling 
selectively removes the lowest entropy gas (see, e.g., Bryan 2000; 
Voit et al.\ 2002), leaving only high entropy (long cooling time) 
gas remaining in the simulated clusters.  However, all the simulations
just mentioned suffer from the overcooling problem (Balogh et al.\ 2001),
so it is not clear to what extent the large entropy cores in clusters
{\it in either mesh or SPH} simulations are produced by shock heating, 
overcooling, or both.  All of these simulations 
implement different prescriptions for radiative cooling (e.g., 
metal-dependent or not), star formation, and feedback, and this may 
lie at the heart of the different findings.  A new generation of 
cosmological code comparisons will be essential in sorting out these 
apparently discrepant findings.  The focus should not only be on 
understanding the differences in the hot gas properties, but also on the 
distribution and amount of stellar matter, as the evolution of the cold 
and hot baryons are obviously intimately linked.  Reasonably tight limits 
on the amount of baryonic mass in the form of stars now exists (see, 
e.g., Balogh et al.z 2008) and provides a useful target for the next 
generation of simulations.  At present, merger-induced mixing 
as an explanation for intrinsic scatter in the hot gas properties 
of groups and clusters remains an open question.
  
Secondly, we have learnt a great deal about the nature of gas 
accretion and the development of hot gas haloes from SPH 
simulations of the universe.  Since we now see that these 
simulations may underestimate the degree of mixing that 
occurs, which of these results are robust, which need 
revision?  For example, Keres et al.\ (2005) (among others) 
have argued that cold accretion by galaxies plays a dominant 
role in fuelling the star formation in galaxies. Is it 
plausible that turbulent eddies could disrupt and mix such 
cold streams as they try to penetrate through the hot halo?  
We can estimate the significance of the effect by comparing 
the Kelvin-Helmholtz timescale, $\tau_{\rm KH}$, with the 
free-fall time, $\tau_{\rm FF}$.  The KH timescale is given 
by (see, e.g., Nulsen 1982; Price 2008)

\begin{equation}
\tau_{\rm KH} \equiv \frac{2 \pi}{\omega}
\end{equation}

\noindent where

\begin{equation}
\omega = \frac{2 \pi}{k} \frac{(\rho \rho')^{1/2} v_{\rm rel}}{(\rho+\rho')}
\end{equation}

\noindent and $\rho$ is the density of the hot halo, $\rho'$ 
is the density of the cold stream, $k$ is the wave number of 
the instability, and $v_{\rm rel}$ the velocity of the stream 
relative to the hot halo.  If the stream and hot halo are in 
approximate pressure equilibrium, this implies a large 
density contrast (e.g., a $10^4$ K stream falling into a 
$10^6$ K hot halo of a Milky Way-type system would imply a 
density contrast of $100$).  In the limit of $\rho' \gg 
\rho$ and recognising that the mode responsible for the 
destruction of the stream is comparable to the size of the 
stream (i.e., $k \sim 2 \pi / r'$), eqns.\ (7) and (8) reduce 
to:

\begin{equation}
\tau_{\rm KH} \approx \frac{r'}{v_{\rm rel}}\biggl(\frac{\rho'}{\rho}\biggr)^{1/2}
\end{equation}

Adopting $\rho'/\rho = 100$, $r' = 100$ kpc, and $v_{\rm rel} 
= 200$ km/s (perhaps typical numbers for a cold stream 
falling into a Milky Way-type system), we find $\tau_{\rm 
KH} \sim 5$ Gyr.  The free-fall time, $\tau_{\rm FF} = R_{\rm 
vir}/V_{\rm circ}(R_{\rm vir})$ [where $R_{\rm vir}$ is the 
virial 
radius of main system and $V_{\rm circ}(R_{\rm vir})$ is the 
circular velocity of the main system at its virial radius], 
is $\sim 1$ Gyr for a Milky Way-type system with mass 
$M_{\rm vir} \sim 10^{12} M_\odot$.  On this basis, it 
seems that the stream would be stable because of the large 
density contrast in the flows.  It is clear, however, that 
the universality of these effects need to be treated with 
caution, as the free-fall and KH timescales are not vastly 
discrepant.  High resolution mesh simulations (cosmological 
or idealised) of Milky Way-like systems would provide a 
valuable check of the SPH results.

Finally, the SPH method has great advantages in terms of 
computational speed, effective resolution and Galilean 
invariance. Is it therefore possible to keep these advantages 
and add additional small scale transport processes to the 
code in order to offset the suppression of mixing?  Wadsley 
et al.\ (2008) and Price (2008) have presented possible 
approaches based on including a thermal diffusion term in 
the SPH equations. Although the approaches differ in their 
mathematical details, the overall effect is the same. 
However, it is not yet clear how well 
this approach will work in cosmological simulations that 
include cooling (and feedback), since the thermal diffusion 
must be carefully controlled to avoid unphysical suppression 
of cooling in hydrostatic regions (e.g., Dolag et al.\ 2005). 
One possibility might be to incorporate such terms as a 
negative surface tension in regions of large entropy 
contrast (Hu \& Adams 2006).  An alternative approach is to 
combine the best features of the SPH method, such as the way 
that it continuously adapts to the local gas density and 
its flow, with the advantage of a Riemann based method of 
solving the fluid dynamic equations (e.g., Inutsuka 2002). 

Clearly, there is a great need to find simple problems in 
which to test these codes: simple shock tube experiments are 
not sufficient because they do not include the disordered 
fluid motions that are responsible for generating the 
entropy core.  Idealised mergers represent a step forward, 
but the problem is still not sufficiently simple that it is 
possible to use self-similar scaling techniques (e.g., 
Bertschinger 1985, 1989) to establish the correct solution . 
One  possibility is to consider the generation of turbulent 
eddies in a fluctuating gravitational potential. We have 
begun such experiments, but (although the fluid flow patterns 
are clearly different) simply passing a gravitational 
potential through a uniform plasma at constant velocity does 
not expose the differences between SPH and Eulerian mesh 
based methods that we see in the idealised merger case. We 
will tackle the minimum complexity that is needed to generate 
these differences in a future paper.

\section*{Acknowledgements}

The authors thank the referee for a careful reading of the manuscript 
and suggestions that improved that paper.  They also thank Volker 
Springel, Mark Voit, and Michael Balogh for very helpful discussions. NLM 
and RAC acknowledge support from STFC studentships.  IGMcC acknowledges 
support from a NSERC Postdoctoral Fellowship at the ICC in Durham and a 
Kavli Institute Fellowship at the Kavli Institute for Cosmology, 
Cambridge. These simulations were performed and analysed on
COSMA, the HPC facilities at the ICC in Durham and we gratefully
acknowledge kind support from Lydia Heck for computing support. The
\flash software used in this work was in part developed by the
DOE-supported ASC / Alliance Center for Astrophysical Thermonuclear
Flashes at the University of Chicago.

\bsp

\label{lastpage}

\end{document}